\documentclass[twocolumn]{aastex631}

\usepackage{xspace}

\newcommand{\R}{\mathcal{R}}
\newcommand{\T}{\mathcal{T}}

\newcommand{\Z}{12+\rm{log\,}(O/H)}
\newcommand{\Zg}{$\nabla[\text{O/H}]$\xspace}
\newcommand{\mstar}{$M_{*}$\xspace}

\usepackage{graphicx,grffile,ulem,url,threeparttable,xcolor,soul,booktabs,natbib,amsmath,mathrsfs}
\usepackage{txfonts,rotating, needspace,subfigure,multirow,array}
\usepackage{enumitem}
\usepackage{float}
\usepackage{placeins}

\newcommand{\rhoTidal}{-0.23}
\newcommand{\pTidal}{0.376}

\newcommand{\rev}[1]{\textcolor{black}{#1}}

\newcommand{\Hb}{\textrm{H}\ensuremath{\beta}\xspace}

\newcommand{\OII}{[\textrm{O}~\textsc{ii}]\xspace}
\newcommand{\OIII}{[\textrm{O}~\textsc{iii}]\xspace}

\newcommand{\NII}{[\textrm{N}~\textsc{ii}]\xspace}

\def\p{{\rm prior}}

\shorttitle{NIRISS PASSAGE - metallicity gradient}
\shortauthors{Li et al.}
\begin{document}

\title{The JWST/PASSAGE Survey: Metallicity Radial Gradients in Emission Line Galaxy Pairs at Cosmic Noon with NIRISS\footnote{Based on observations acquired by the JWST under the GO program ID 1571 (PI M. Malkan)}}

\correspondingauthor{Xin Wang}
\email{xwang@ucas.ac.cn}


\author[0000-0003-4813-8482]{Sijia Li}
\affil{Department of Astronomy, School of Physics and Astronomy, Shanghai Jiao Tong University, 800 Dongchuan Road, Shanghai 200240, China}
\affil{School of Astronomy and Space Science, University of Chinese Academy of Sciences (UCAS), Beijing 100049, China}
\affil{Department of Astronomy, Xiamen University, Xiamen, Fujian 361005, China}

\author[0000-0002-9373-3865]{Xin Wang}
\affil{School of Astronomy and Space Science, University of Chinese Academy of Sciences (UCAS), Beijing 100049, China}
\affil{National Astronomical Observatories, Chinese Academy of Sciences, Beijing 100101, China}
\affil{Institute for Frontiers in Astronomy and Astrophysics, Beijing Normal University,  Beijing 102206, China}

\author[0000-0001-6919-1237]{Matthew A. Malkan}
\affiliation{Department of Physics and Astronomy, University of California, Los Angeles, 430 Portola Plaza, Los Angeles, CA 90095, USA}

\author[0000-0001-5860-3419]{Tucker Jones}
\affiliation{Department of Physics and Astronomy, University of California Davis, 1 Shields Avenue, Davis, CA 95616, USA}

\author[0000-0001-7166-6035]{Vihang Mehta}
\affil{Infrared Processing and Analysis Center, Caltech, 1200 E. California Blvd., Pasadena, CA 91125, USA}

\author[0000-0002-7928-416X]{Y. Sophia Dai}
\affil{National Astronomical Observatories, Chinese Academy of Sciences, Beijing 100101, China}

\author[0000-0001-8587-218X]{Matthew J. Hayes}
\affiliation{Stockholm University, Department of Astronomy and Oskar Klein Centre for Cosmoparticle Physics, AlbaNova University Centre, SE-10691, Stockholm, Sweden.}

\author[0000-0002-9136-8876]{Claudia Scarlata}
\affiliation{Minnesota Institute for Astrophysics, University of Minnesota, Twin Cities, 116 Church St SE, Minneapolis, MN 55455, USA}

\author[0000-0002-7064-5424]{Harry I. Teplitz}
\affil{Infrared Processing and Analysis Center, Caltech, 1200 E. California Blvd., Pasadena, CA 91125, USA}

\author[0000-0003-0980-1499]{Benedetta Vulcani}
\affiliation{INAF, Osservatorio Astronomico di Padova, Vicolo dell'Osservatorio 5, 35122 Padova, Italy}


\author[0000-0003-4804-7142]{Ayan Acharyya}
\affiliation{INAF -- Osservatorio Astronomico di Padova, Vicolo Osservatorio 5, 35122 Padova, Italy}

\author[0000-0002-7570-0824]{Hakim Atek}
\affiliation{CNRS, Institut d'Astrophysique de Paris, 98 bis Boulevard Arago, 75014 Paris, France}

\author[0000-0002-8651-9879]{Andrew J Bunker}
\affiliation{Department of Physics, University of Oxford, Denys Wilkinson Building, Keble Road, Oxford OX13RH, U.K.}

\author[0000-0002-0072-0281]{Farhanul Hasan}
\affiliation{Space Telescope Science Institute, 3700 San Martin Drive, Baltimore, MD, 21218 USA}

\author[0009-0002-9932-4461]{Mason S. Huberty}
\affiliation{Minnesota Institute for Astrophysics, University of Minnesota, Twin Cities, 116 Church St SE, Minneapolis, MN 55455, USA}

\author[0000-0002-0364-1159]{Zahra Sattari}
\affiliation{IPAC, California Institute of Technology, 1200 E. California Blvd, Pasadena, CA 91125, USA}

\author[0000-0003-3108-0624]{Peter J. Watson}
\affiliation{INAF -- Osservatorio Astronomico di Padova, Vicolo Osservatorio 5, 35122 Padova, Italy}

\newcommand{\approptoinn}[2]{\mathrel{\vcenter{
  \offinterlineskip\halign{\hfil$##$\cr
    #1\propto\cr\noalign{\kern2pt}#1\sim\cr\noalign{\kern-2pt}}}}}

\newcommand{\appropto}{\mathpalette\approptoinn\relax}

\begin{abstract}

Using \textit{JWST}/NIRISS slitless spectroscopy from the PASSAGE program, we analyze 17 spectroscopically confirmed emission-line galaxies in candidate interacting systems at $z \sim 1.7\text{--}3.2$.
These systems span stellar masses of $7.3 \lesssim \log(M_*/M_\odot) \lesssim 10.4$ and global gas-phase metallicities of $12 + \log(\text{O/H}) \sim 7.6\text{--}8.7$. From spatially resolved emission line maps, we measure their radial metallicity gradients separately for each galaxy, which are in the range $\nabla[\text{O/H}] \sim -0.2$ to $+0.45~\text{dex kpc}^{-1}$. 
The mean metallicity gradients in bins of stellar mass and redshift are consistent with previous measurements, and their scatter is not significantly different from that of the heterogeneous literature compilation in either the $\nabla[\text{O/H}]$--$M_*$ or $\nabla[\text{O/H}]$--$z$ relation.
Interestingly, we find tentative evidence for an anti-correlation between $\nabla[\text{O/H}]$ and the metallicity offset from the mass--metallicity relation, $\nabla_{\rm [O/H]} =(-0.30\pm0.06)\,\Delta[12+\log(\mathrm{O/H})]_{\rm obs-MZR(z,\,M_{*})} + 0.04$, possibly reflecting interaction-driven variations in star formation and chemical enrichment.
Moreover, galaxies in pairs with comparable component masses tend to exhibit positive $\nabla[\text{O/H}]$, whereas galaxies in unequal-mass pairs tend to exhibit negative $\nabla[\text{O/H}]$; neither trend is statistically significant in the present sample. Overall, our results reveal substantial diversity in the metallicity distributions of candidate interacting systems at cosmic noon and provide tentative evidence that galaxy interactions may influence the redistribution of gas and metals.
\end{abstract}


    \section{Introduction}  \label{sec:intro}
    The epoch known as “cosmic noon” ($z \sim 1$--3) marks a peak in the cosmic star formation rate and a heightened frequency of galaxy mergers \citep{2011Lotz,2014Madau}. This period represents a crucial phase in galaxy evolution, during which gravitational interactions and gas dynamics actively shape the internal structure and chemical composition of galaxies.

    Gas-phase metallicity, traced by emission lines from the ionized interstellar medium (ISM), encodes the history of star formation and gas flows such as accretion, feedback-driven outflows, and gas recycling \citep{2019Maiolino, Shin2021, 2025Curti}, while recent studies further show that the physical conditions of ionized gas, such as electron density, are closely linked to star formation activity \citep{LSJ2025b}. Gas-phase metallicity is commonly expressed as $12+\log(\mathrm{O/H})$, which represents the logarithmic abundance of oxygen relative to hydrogen, normalized such that the solar value is approximately 8.69 \citep{Asplund2009}. The spatial distribution of metallicity, which is typically quantified by metallicity gradients, provides further insight into galaxy assembly \citep{2015Jones,2017xwang,2019XWang, Barisic2025}. In the local Universe, disk galaxies often exhibit negative radial metallicity gradients, interpreted as a consequence of inside-out growth \citep{ Sanchez2014_CalifaGradient,2015Ho, 2025Chamorro-Cazorla}. However, galaxy mergers can dramatically alter these gradients through gravitational torques and large-scale gas redistribution, often flattening or even inverting them \citep{2010Rupke, 2012Torrey, 2020Bustamante}. Both observations \citep{2010Kewley} and simulations \citep{2010Montuori, 2018Bustamante} support the idea that interactions trigger gas mixing and dynamically reshape a galaxy’s chemical profile.

    Systematic measurements of metallicity gradients in interacting systems across cosmic time are thus critical for constraining models of gas inflows, star formation regulation, and the role of mergers in galaxy evolution \citep{2010Kewley}. Until recently, such studies at high redshift have been limited by the sensitivity and resolution of available instruments.

    With its unprecedented infrared sensitivity and resolution, the James Webb Space Telescope (JWST) is driving major advances in our understanding of the chemical properties of high-redshift galaxies through both imaging \citep[e.g.,][]{Treu2022, Kartaltepe2023, Xu2024, Yu2025, Yu2026b, Shuntov2025} and spectroscopy \citep[e.g.,][]{Sanders2024, Backhaus2024, Fujimoto2024, Schouws2024, 2025Ju,LSJ2025}. In particular, the Near Infrared Imager and Slitless Spectrograph (NIRISS) provides slitless spectroscopy over the 0.8–2.2 $\mu$m range, enabling the simultaneous detection of key rest-frame optical emission lines (e.g., [O\,\textsc{ii}], H$\beta$, [O\,\textsc{iii}], and [S\,\textsc{ii}]) for galaxies at $z > 1$. These lines enable gas-phase metallicity and star-formation measurements at high redshift \citep{2022XWang_glass,2024Venturi,li202513billionyearviewgalaxygrowth,2026Acharyya}. JWST/NIRCam imaging also reveals detailed galaxy morphologies, which are especially valuable for identifying and characterizing candidate interactions.

    Enabled by JWST’s capabilities, this study draws on data from the JWST/NIRISS Parallel Application of Slitless Spectroscopy to Analyze Galaxy Evolution (PASSAGE; \citealt{2025Malkan}), a NIRISS wide-field slitless spectroscopy program that samples high-redshift galaxies in parallel with deep imaging.
    We focus on morphologically disturbed and interacting galaxies at $z \sim 2-3$, with the aim of measuring spatially resolved gas-phase metallicity gradients. 
    Our goal is to examine how interaction-driven processes influence chemical enrichment patterns, and how these internal gradients correlate with global galaxy properties such as stellar mass.
    By probing the internal structure of galaxies in candidate interacting systems at cosmic noon, this work provides new constraints on the interplay between dynamical evolution and chemical enrichment in the early universe.

    This paper is organized as follows. Section~\ref{sec:datsam} describes the dataset and sample selection. Section~\ref{sec:analysis} presents our analysis and gradient measurements. In Section~\ref{sec:rslt}, we discuss the implications of our results. In Section~\ref{sec:conclu}, we list our conclusions. In this work, we adopt the \textit{Planck} 2018 cosmology: $H_0 = 67.4$ km s$^{-1}$ Mpc$^{-1}$, $\Omega_m = 0.315$, and $\Omega_\Lambda = 0.685$ \citep{2020Plank18}.

\section{Spectroscopic and Imaging Data} \label{sec:datsam}
The observations used in this work include two types of data: galaxy imaging from the COSMOS \citep{2007ApJS..172..196K,2011ApJS..197...35G} and COSMOS-Web survey \citep{2023Casey} and slitless spectroscopy from the PASSAGE survey \citep{2025Malkan}. We use the imaging data to perform photometry, while the slitless spectroscopy is used to measure redshifts and derive metallicities.

\subsection{Spectroscopy Data}

The PASSAGE survey comprises 63 parallel NIRISS slitless spectroscopic fields. We focus on the Parallel-28 field, which provides high-quality data and complete three-filter NIRISS coverage within the COSMOS region observed at two orientations, enabling reliable redshift determination and emission-line--based metallicity measurements.
The PASSAGE Parallel-28 field lies within the well-studied COSMOS field \citep{2007cosmos}, which benefits from extensive multi-wavelength observations.
In the same manner as \cite{2025Runnholm}, the NIRISS data used in this work were retrieved from the Mikulski Archive for Space Telescopes (MAST). The raw exposures were first processed through the standard JWST calibration pipeline to produce rate files.
\rev{The spectral extraction and contamination modeling follow the PASSAGE reduction described by \citet{Huberty2026}. We used the updated PASSAGE products generated with Grizli version 1.12.11 \citep{2019Brammer,2023Brammer}. Grizli models the dispersed spectra of all sources identified in the direct-image segmentation map in each exposure, thereby accounting simultaneously for contamination from neighboring objects before producing the one- and two-dimensional spectra. Combined spectra were generated where both GR150R and GR150C observations were available.}
These two grisms disperse the wavelengths along the detector rows and columns, respectively, allowing us to separate features from overlapping spectra.

Final image combination and alignment were performed using the AstroDrizzle routine from the DrizzlePac software package \citep{2012Gonzaga, 2021Hoffmann}, resulting in distortion-corrected and background-subtracted frames suitable for spectroscopic analysis.

\subsection{Imaging Data}

We use JWST data from the COSMOS-Web survey. The COSMOS-Web survey delivers the largest contiguous extragalactic imaging dataset from JWST to date, covering a 0.54 deg$^2$ region of the COSMOS field with NIRCam in four filters: F115W, F150W, F277W, and F444W. These bands sample the rest-frame UV to optical light for galaxies at a wide range of redshifts, enabling detailed studies of galaxy morphology, stellar mass distribution, and recent star formation. 

The resulting imaging achieves 5$\sigma$ depths ranging from 26.7 to 28.3 AB magnitudes (depending on filter and region), with the deepest exposures in the central field (more detailed see in \citealt{Franco2025cosmos_web}).

The imaging data used in this work are taken from COSMOS-Web Data Release 1 (DR1)\footnote{\url{https://cosmos2025.iap.fr/}}. 
The raw data were processed using the standard JWST Calibration Pipeline (v1.11.2 \citealt{bushouse_2023_7577320}) for detector-level corrections, flat-fielding, and photometric calibration (see details in \citealt{Franco2025cosmos_web}). Additional custom processing was performed to improve background subtraction, identify and mask bad pixels, and refine astrometric alignment using external Gaia DR3 and HST reference catalogs. Mosaics were created using ResampleStep with a pixel scale of 30 mas pixel$^{-1}$. We also make use of HST/ACS F814W imaging from the CANDELS survey \citep{2011ApJS..197...35G}.

\begin{figure*}
        \centering

    \includegraphics[width=0.9\linewidth]{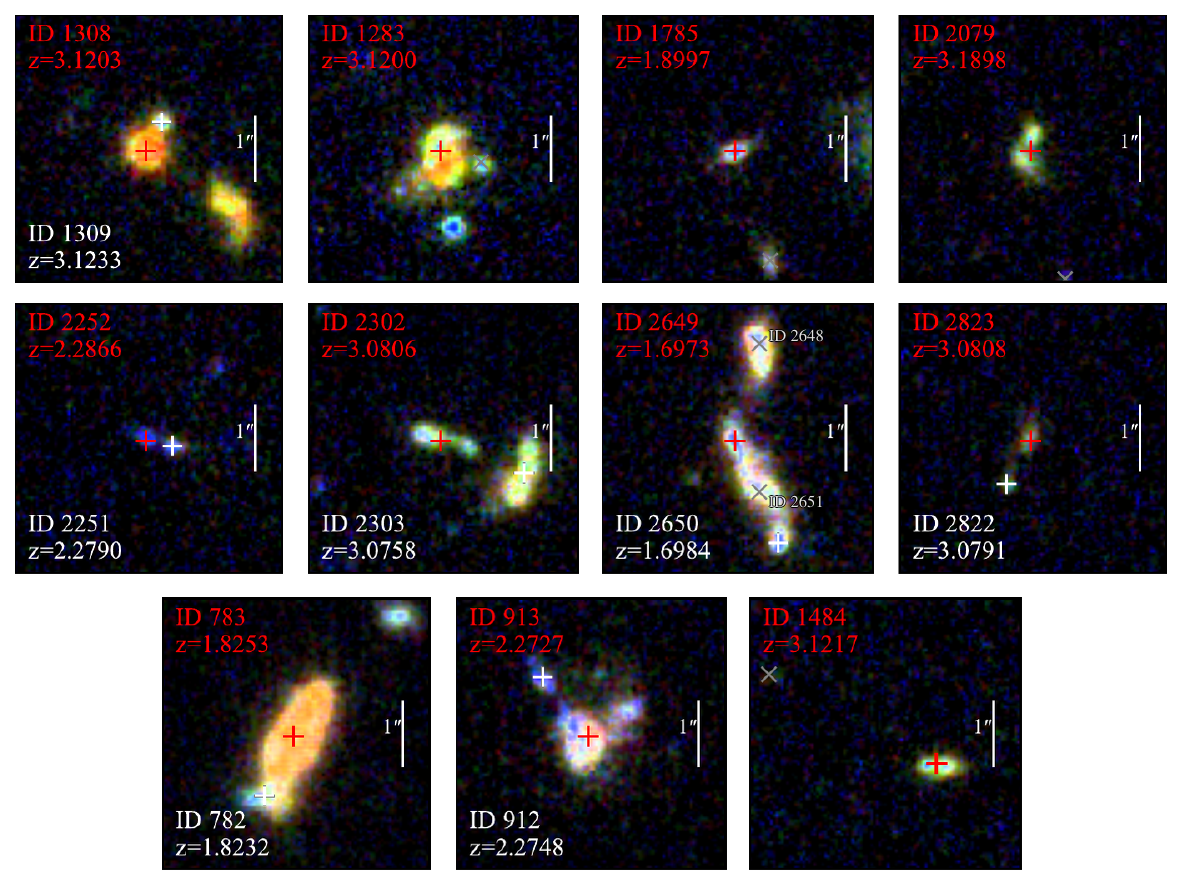}
    \caption{RGB color-composite images of spectroscopically confirmed emission-line galaxy pairs identified in the PASSAGE Par028 field. 
    Each panel shows a galaxy group identified as an interaction pair likely in merger, based on projected proximity, disturbed morphology, and redshift similarity. 
    The RGB images are constructed using NIRCam filters F150W (blue), F277W (green), and F444W (red). 
    Each stamp is $4\arcsec \times 4\arcsec$ in size, with a white bar indicating $1\arcsec$. 
    Galaxy IDs and redshifts are labeled in the top and bottom corners of each panel, respectively. 
    The galaxy centers used for gradient measurements are marked in red and white, with labels in the corresponding colors, and are only defined for galaxies with available metallicity gradient measurements. Galaxy centers shown in grey are also group members but lack reliable metallicity measurements. The cutout containing Galaxies 2648--2651 shows a single four-member interacting system, which is represented by two pairwise entries in Table~\ref{tab:1}.}

    \label{fig:rgb}
\end{figure*}
\subsection{Sample Selection: Identifying Galaxy Pairs}

To study the metallicity and gas dynamics in interacting galaxies, we focus on galaxy pairs likely undergoing interactions. A critical step is to identify galaxy pairs that are likely in the process of merging or have recently interacted  \citep{2021Dai}. We select interacting pairs based on the following criteria, adapted from observational studies \citep[e.g.,][]{Ellison2008,Patton2013}:

\begin{enumerate}[label=(\alph*)]
    \item \textbf{Projected Separation:} The two galaxies must have a projected distance $R_\mathrm{proj} \lesssim 50$ kpc. This threshold captures both pre-merger and coalescing pairs and is commonly used in the literature to identify strongly interacting systems \citep{Patton2013}.

    \item \textbf{Morphological Disturbance:} At least one member of the pair must show signs of tidal features in their morphology, such as tails, bridges, or asymmetries, suggesting a gravitational interaction. 
    We perform visual inspection of the high-resolution JWST/NIRCam F444W images to identify such features.

    \item \textbf{  Spectroscopic-redshift consistency :}   Both members must have spectroscopic redshifts measured from the NIRISS spectra. We adopt the Grizli $z_{50}$ estimates reported in the PASSAGE Par028 spectroscopic catalog (\texttt{Par028\_speccat.fits}), produced as part of the spectroscopic reduction. We require  $\Delta z \lesssim 0.01$   (approximately $\Delta v_{\rm los}\lesssim700$ km s$^{-1}$ over the redshift range considered). This inclusive threshold retains plausible associations given the redshift uncertainties; a more conservative $500$ km s$^{-1}$ cut would reduce contamination but also remove candidates whose uncertainties overlap that boundary .

    \item \textbf{Redshift requirement:} $1.6 \lesssim z \lesssim 3.2$, such that the key rest-frame optical emission lines used for our metallicity measurements (\OII, H$\beta$, and \OIII) fall within the NIRISS wavelength range and can be measured with sufficient SNR.

\end{enumerate}

We further select galaxies for metallicity measurements based on the following signal-to-noise ratio (SNR) thresholds: $\rm SNR(\OIII\, \lambda5007) > 5$, $\rm SNR(\OII) > 1$, and $\rm SNR(\Hb) > 2$.   In addition to these individual-line cuts, the Bayesian likelihood in Section~\ref{subsec:M} propagates the flux uncertainties of all lines entering each ratio. Thus, a low-SNR \OII\ or \Hb\  measurement produces a correspondingly broad metallicity posterior rather than being treated as a precise ratio.

We  cross-match   the sample against the X-ray-selected COSMOS catalog of \citet{2016Marchesi} and find no counterparts. Because that catalog is defined by  X-ray  detection , this non-detection cannot exclude Compton-thick or intrinsically faint active galactic nuclei (AGN). A spatially resolved BPT analysis is not possible for the full sample because H$\alpha$ and \NII\ fall outside the NIRISS wavelength coverage for the higher-redshift systems and are blended at the NIRISS grism spectral resolution ($R\sim150$) where covered. As an independent check, we construct a mass--excitation (MEx) diagram using $\log([\mathrm{O\,III}]\lambda5007/\mathrm{H}\beta)$\ from the summed spatial-bin line fluxes and the stellar masses derived from our SED fitting (Section~\ref{subsec:sed}). All galaxies lie below the MEx AGN boundary after applying the $+0.75$~dex\ stellar-mass shift proposed for $z\sim2.3$\ samples \citep{Juneau2014,Coil2015}, and hence this test identifies no clear AGN candidates. Because the MEx diagnostic is not securely calibrated over the full redshift range of our sample, particularly at $z\gtrsim3$, residual AGN contamination remains a systematic limitation .
 All selected galaxies, along with their interacting companions, are listed in Table~\ref{tab:1}, and their RGB images are shown in Figure~\ref{fig:rgb}.

The final quantitative sample is a subset of the galaxies listed in Table~\ref{tab:1} and comprises 17 galaxies with reliable metallicity-gradient measurements, because not every galaxy has a sufficient number of spatial elements with adequate SNR. Galaxy~1282 is shown for completeness but is not included in the quantitative analysis because its two usable spatial bins lie at nearly identical galactocentric radii, leaving its gradient unconstrained.
These galaxy pairs are physically close and   have small line-of-sight velocity differences, but projected proximity does not uniquely establish that a system is gravitationally bound: some candidates may be unbound fly-bys \citep{Moreno2013}. Without resolved kinematics, neither the orbital geometry nor the phase relative to pericentric passage is known. Moreover, compact star-forming clumps, which are common in galaxies near cosmic noon, can mimic companions \citep{ForsterSchreiber2011,Shibuya2016,Kalita2025}. We mitigate this by requiring a separately identified source with its own spectroscopic redshift and by inspecting the multi-band morphology, although these checks cannot eliminate every ambiguous case. Mixing unbound systems and different interaction stages into the pair sample would tend to dilute intrinsic trends.

\begin{table*}[htbp]
  \centering
  \caption{Observed properties of our spectroscopically confirmed emission-line galaxy pairs.}
  \label{tab:1}
  \begin{tabular}{cccc|cccc|c}
\toprule
$\rm ID_1$ & R.A. & Dec. & $z$ & $\rm ID_2 $& R.A. & Dec. & $z$ & Distance \\
&  (deg)& (deg)&&&(deg)&(deg)&&(kpc) \\
(1)&(2)&(3)&(4)&(5)&(6)&(7)&(8)&(9)\\

\midrule

782  & 150.084256 & 2.409560 & $1.8233$ & 783  & 150.084231 & 2.409766 & $1.8252$&      6.5 \\  
912  & 150.089224 & 2.411302 & $2.2746$ & 913  & 150.089034 & 2.411205 & $2.2724$&      6.5 \\  
1282 & 150.081781 & 2.415710 & $3.1198$ & 1283 & 150.081916 & 2.415689 & $3.1201$&      3.8 \\  
1308 & 150.083155 & 2.415907 & $3.1206$ & 1309 & 150.083125 & 2.416017 & $3.1231$&      3.2 \\  
1484 & 150.103385 & 2.417769 & $3.1216$ & 1485 & 150.103966 & 2.417847 & $3.1224$&      16.4 \\ 
1765 & 150.090584 & 2.421075 & $1.9023$ & 1785 & 150.090772 & 2.421361 & $1.8995$&      10.6 \\ 
2044 & 150.101006 & 2.424405 & $3.1905$ & 2079 & 150.101248 & 2.424775 & $3.1902$&      12.3 \\ 
2251 & 150.080179 & 2.426842 & $2.2790$ & 2252 & 150.080257 & 2.426862 & $2.2866$&      2.4 \\  
2302 & 150.079836 & 2.427399 & $3.0801$ & 2303 & 150.079542 & 2.427353 & $3.0757$&      8.4 \\  
2648 & 150.098781 & 2.431708 & $1.6960$ & 2650 & 150.098915 & 2.431777 & $1.6984$&      5.5 \\  
2651 & 150.098678 & 2.431566 & $1.6973$ & 2649 & 150.098947 & 2.432087 & $1.6973$&      9.7 \\  
2822 & 150.080063 & 2.434127 & $3.0796$ & 2823 & 150.080032 & 2.434301 & $3.0809$&      5.0 \\  

\bottomrule
\end{tabular}

\tablecomments{
For each row, Cols.~(1)--(4) and Cols.~(5)--(8) give the properties of the two galaxies in one candidate interacting system.
Col.~(1); galaxy ID in PASSAGE parallel 28 field. Col.~(2): right ascension in J2000. Col.~(3): declination in J2000. Col.~(4): spectroscopic redshift. Col.~(5-8) is similar with Col.~(1-4). Col.~(9): projection distance between galaxy $\rm ID_1$ and galaxy $\rm ID_{2}$.}

\end{table*}

\begin{table*}[htbp]
    \centering
    \caption{Derived physical properties of individual galaxies in our sample of interacting galaxy pairs.}
    \label{tab:2}
    \resizebox{\textwidth}{!}{%
\begin{tabular}{cccccccccc}
\toprule
ID & $z_{\rm spec}$ & $n_{\rm s}$ & $R_e$ & $12+\log({\rm O/H})$ & $\log(M_*/M_\odot)$ & \Zg & $A_V$ & SFR & Pair mass ratio $\mu$ (primary ID) \\
 & & & (arcsec; kpc) & & & (dex kpc$^{-1}$; dex $R_e^{-1}$) & & ($M_\odot$ yr$^{-1}$) & \\
\midrule
782  &$1.8233$&$0.74$&$0.26; 2.25$&$8.69_{-0.05}^{+0.04}$&$8.88_{-0.08}^{+0.07}$&$-0.12\pm0.12; -0.27\pm0.27$&$0.30_{-0.22}^{+0.39}$&$3.78_{-2.90}^{+2.04}$&$0.03$ (783)\\
783  &$1.8252$&$1.21$&$0.43; 3.72$&$8.70_{-0.04}^{+0.03}$&$10.38_{-0.08}^{+0.06}$&$0.01\pm0.03; 0.04\pm0.11$&$0.33_{-0.23}^{+0.33}$&$8.24_{-6.40}^{+3.65}$&\\
912  &$2.2746$&$0.89$&$0.15; 1.26$&$8.28_{-0.05}^{+0.05}$&$8.23_{-0.11}^{+0.16}$&$-0.03\pm0.04; -0.04\pm0.05$&$0.40_{-0.28}^{+0.37}$&$6.48_{-4.70}^{+3.51}$&$0.03$ (913)\\
913  &$2.2724$&$1.34$&$0.22; 1.85$&$8.44_{-0.05}^{+0.04}$&$9.83_{-0.07}^{+0.04}$&$0.05\pm0.01; 0.09\pm0.02$&$0.66_{-0.30}^{+0.31}$&$37.15_{-26.62}^{+15.38}$&\\
1282$^{\dagger}$ &$3.1198$&$1.29$&$0.08; 0.62$&$8.54_{-0.07}^{+0.06}$&$7.79_{-0.09}^{+0.16}$&$-0.05\pm6.59; -0.03\pm4.11$&$1.11_{-0.50}^{+0.52}$&$14.69_{-7.96}^{+12.77}$&$0.04$ (1283)\\
1283 &$3.1201$&$3.05$&$0.20; 1.56$&$8.36_{-0.05}^{+0.05}$&$9.22_{-0.06}^{+0.11}$&$0.01\pm0.01; 0.02\pm0.02$&$1.00_{-0.33}^{+0.30}$&$67.24_{-45.95}^{+28.44}$&\\
1308 &$3.1206$&$1.07$&$0.10; 0.78$&$8.43_{-0.07}^{+0.06}$&$9.80_{-0.15}^{+0.17}$&$0.04\pm0.07; 0.03\pm0.05$&$0.44_{-0.30}^{+0.43}$&$9.40_{-6.47}^{+6.59}$&$0.07$ (1308)\\
1309 &$3.1231$&$2.12$&$0.11; 0.86$&$8.39_{-0.09}^{+0.07}$&$8.67_{-0.10}^{+0.15}$&$0.11\pm0.15; 0.09\pm0.13$&$1.50_{-0.58}^{+0.53}$&$27.09_{-13.38}^{+23.85}$&\\
1484 &$3.1216$&$0.36$&$0.16; 1.25$&$8.22_{-0.07}^{+0.07}$&$8.80_{-0.07}^{+0.09}$&$-0.08\pm0.03; -0.10\pm0.04$&$1.11_{-0.55}^{+0.56}$&$27.52_{-14.59}^{+25.28}$&$0.01$ (1485)\\
1785 &$1.8995$&$1.77$&$0.15; 1.29$&$7.63_{-0.15}^{+0.20}$&$8.84_{-0.05}^{+0.08}$&$0.45\pm0.15; 0.58\pm0.19$&$1.34_{-0.96}^{+1.51}$&$3.84_{-1.43}^{+15.06}$&$0.67$ (1785)\\
2079 &$3.1902$&$0.86$&$0.11; 0.85$&$8.12_{-0.06}^{+0.07}$&$9.25_{-0.15}^{+0.06}$&$0.08\pm0.03; 0.07\pm0.03$&$0.26_{-0.19}^{+0.46}$&$11.69_{-9.36}^{+7.85}$&$0.01$ (2079)\\
2251 &$2.2790$&$1.52$&$0.19; 1.60$&$8.36_{-0.09}^{+0.08}$&$8.36_{-0.04}^{+0.04}$&$-0.20\pm0.06; -0.32\pm0.10$&$0.90_{-0.54}^{+0.61}$&$4.50_{-2.29}^{+5.19}$&$0.09$ (2251)\\
2252 &$2.2866$&$0.80$&$0.08; 0.67$&$8.14_{-0.11}^{+0.10}$&$7.34_{-0.03}^{+0.04}$&$0.19\pm0.11; 0.13\pm0.07$&$2.12_{-1.01}^{+0.79}$&$27.82_{-9.12}^{+39.63}$&\\
2302 &$3.0801$&$0.81$&$0.15; 1.17$&$8.25_{-0.05}^{+0.05}$&$9.10_{-0.14}^{+0.17}$&$-0.01\pm0.01; -0.01\pm0.01$&$0.45_{-0.30}^{+0.40}$&$19.78_{-13.89}^{+12.13}$&$0.20$ (2303)\\
2303 &$3.0757$&$0.42$&$0.32; 2.50$&$8.38_{-0.06}^{+0.06}$&$9.77_{-0.05}^{+0.03}$&$0.09\pm0.02; 0.23\pm0.05$&$0.43_{-0.29}^{+0.38}$&$31.25_{-22.06}^{+19.10}$&\\
2649 &$1.6973$&$0.97$&$0.22; 1.91$&$8.35_{-0.13}^{+0.11}$&$8.87_{-0.07}^{+0.06}$&$0.08\pm0.04; 0.15\pm0.08$&$1.48_{-0.87}^{+1.04}$&$15.84_{-5.31}^{+41.26}$&$0.30$ (2651)\\
2650 &$1.6984$&$0.97$&$0.13; 1.13$&$8.31_{-0.41}^{+0.23}$&$8.58_{-0.05}^{+0.05}$&$-0.16\pm0.08; -0.18\pm0.09$&$2.02_{-1.75}^{+1.45}$&$33.77_{-3.93}^{+199.6}$&$0.10$ (2648)\\
2823 &$3.0809$&$1.68$&$0.25; 1.96$&$8.36_{-0.13}^{+0.10}$&$8.97_{-0.12}^{+0.08}$&$-0.13\pm0.18; -0.25\pm0.35$&$1.63_{-0.96}^{+0.81}$&$24.55_{-7.46}^{+41.85}$&$0.12$ (2823)\\
\bottomrule
\end{tabular}}

\tablecomments{The physical radii use the cosmology adopted in this work. The second gradient entry is $\nabla[\mathrm{O/H}]\,R_e$. The pair ratio is $\mu=M_{*,\rm secondary}/M_{*,\rm primary}\leq1$ and is listed once when both members appear; the primary ID is given in parentheses. $^{\dagger}$Galaxy~1282 is shown for completeness but is excluded from all quantitative analyses because its two usable spatial bins lie at nearly identical radii, leaving the gradient unconstrained.}

\end{table*}

\section{Analysis and Measurements}  \label{sec:analysis}

\subsection{Morphology and Photometry Measurement} \label{subsec:morpho}

Interacting galaxies often exhibit complex morphological features, such as tidal tails, double cores, or multiple components within a single system. To robustly characterize their structures and obtain consistent photometric measurements, we perform forward modeling of the galaxy light distributions.

Because many galaxies in our sample are in close proximity, all components within each system are fitted simultaneously to self-consistently account for mutual contamination in their outer light profiles. We adopt hybrid point spread functions (PSFs) from \cite{2026ApJS..283...35Y}, which combine an empirical core with theoretical outer profiles, ensuring accurate modeling of both central structures and extended wings.

\begin{figure*}
    \centering
    \includegraphics[width=0.9\linewidth]{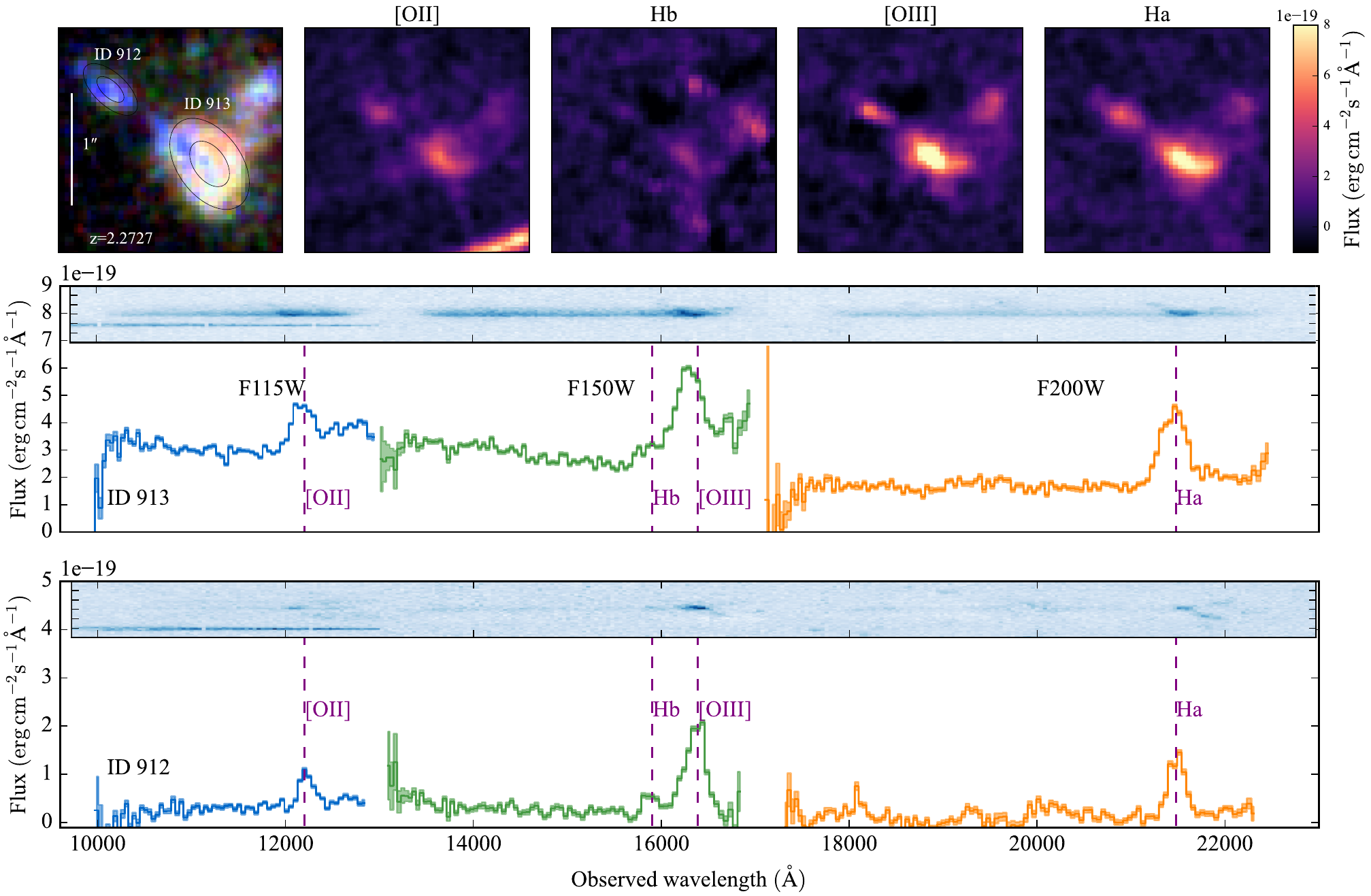}
    \caption{Imaging and spectroscopic data for an interacting galaxy group. The top row presents an RGB composite image (constructed from JWST F150W, F277W, and F444W filters) alongside 2D emission-line maps of [O II], H$\beta$, [O III], and H$\alpha$ (left to right). The second and third rows display the 1D and 2D spectra (both in observer frame) of Galaxy 912 and 913, respectively. Purple dashed lines in the spectra indicate emission lines used for metallicity measurements.}
    \label{fig:1} 
\end{figure*}

\begin{figure*}[!t]
    \centering
      \includegraphics[width=1\linewidth]{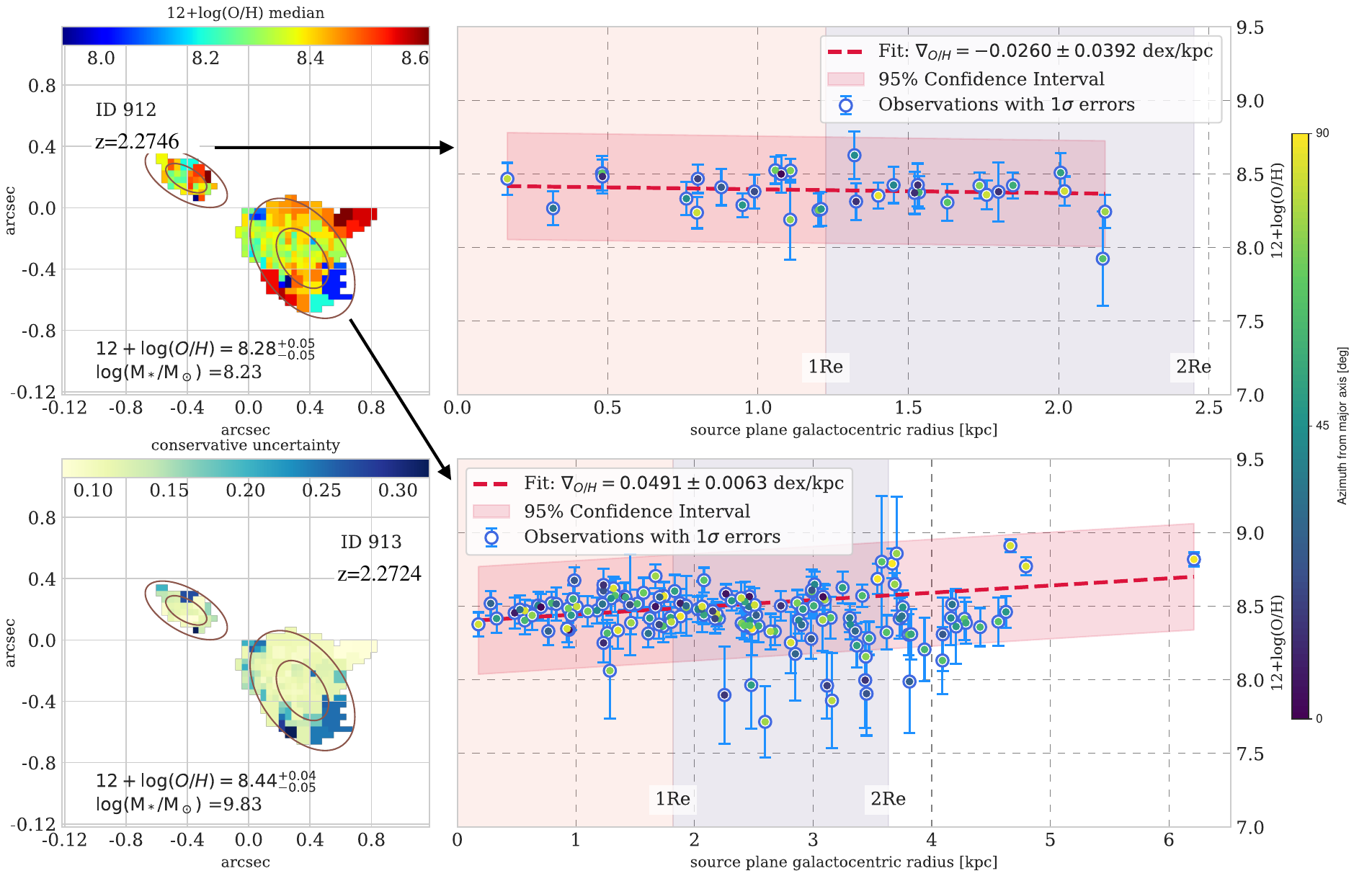}
     \caption{Spatially resolved metallicity and radial profiles for Galaxy Group   912--913 . The left column presents the metallicity maps of the selected galaxies (top) and their associated conservative uncertainties (bottom). The right column shows the radial metallicity profiles for Galaxy 912 (top) and Galaxy 913 (bottom). The  points are color-coded by the deprojected azimuthal angle $\phi$ of each Voronoi-bin centroid, measured from the photometric major axis: $\phi=0^\circ$ corresponds to the major axis and $\phi=90^\circ$ to the minor axis. The  red dashed line shows the linear fitting result, while the red shaded region represents the 95\% confidence interval of the fit. The orange- and blue-shaded regions indicate the areas within 1$R_e$ and 2$R_e$, respectively.}
    \label{fig:2} 
\end{figure*}

\begin{figure}
    \centering
      \includegraphics[width=1\linewidth]{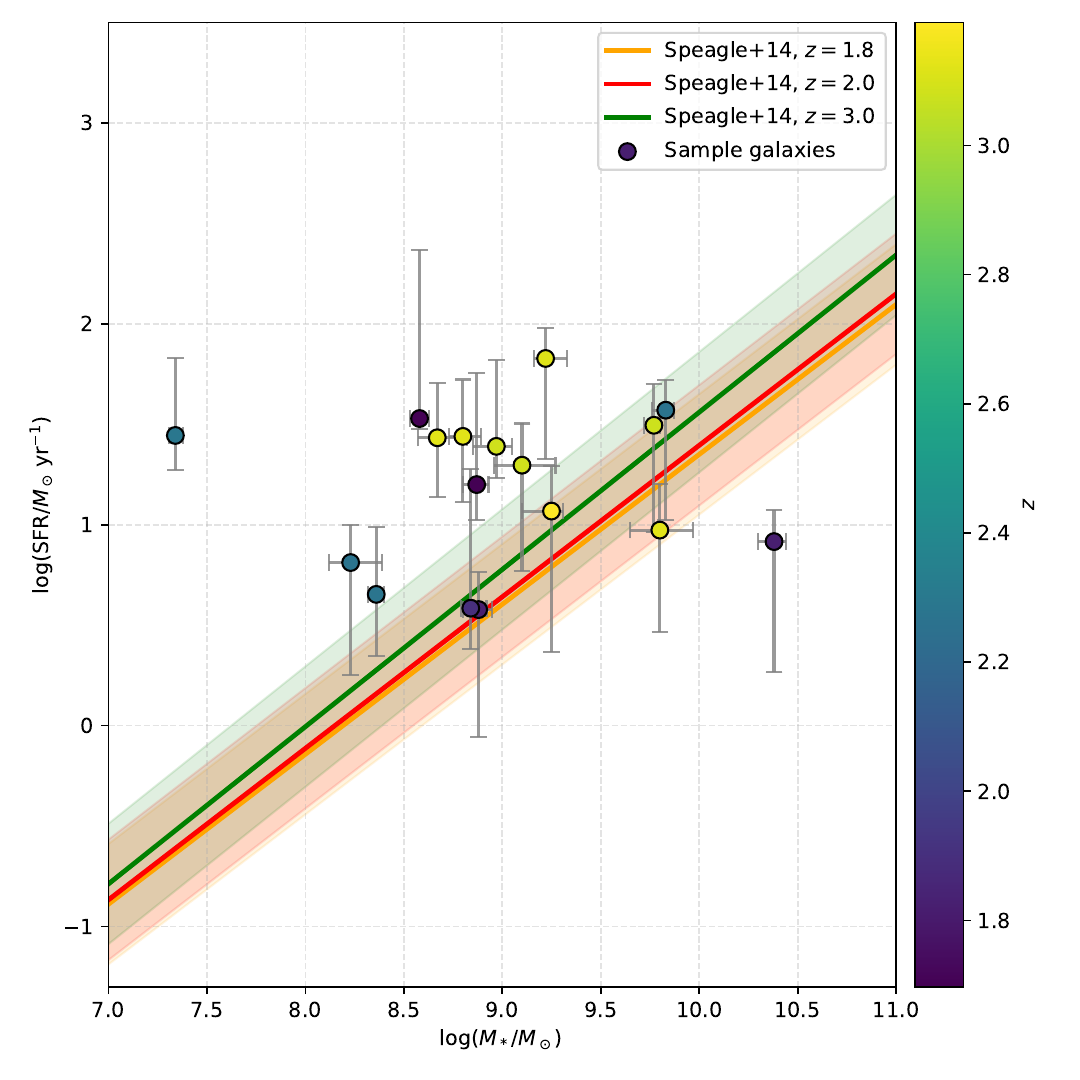}
     \caption{Star formation rate (SFR) as a function of stellar mass ($M_*$). The   orange , red ,  and green lines show the   star-forming main-sequence relations  at $z=1.8$, 2.0 ,  and 3.0  , respectively, from  \citet{Speagle2014}  ; the  shaded   regions indicate their $3\sigma$ uncertainties .   The sample galaxies are shown  as  circles with grey error bars  and are color-coded by their spectroscopic redshifts, as indicated by the color bar .}
    \label{fig:SFMS} 
\end{figure}

\begin{figure*}
    \centering
    \includegraphics[width=1\linewidth]{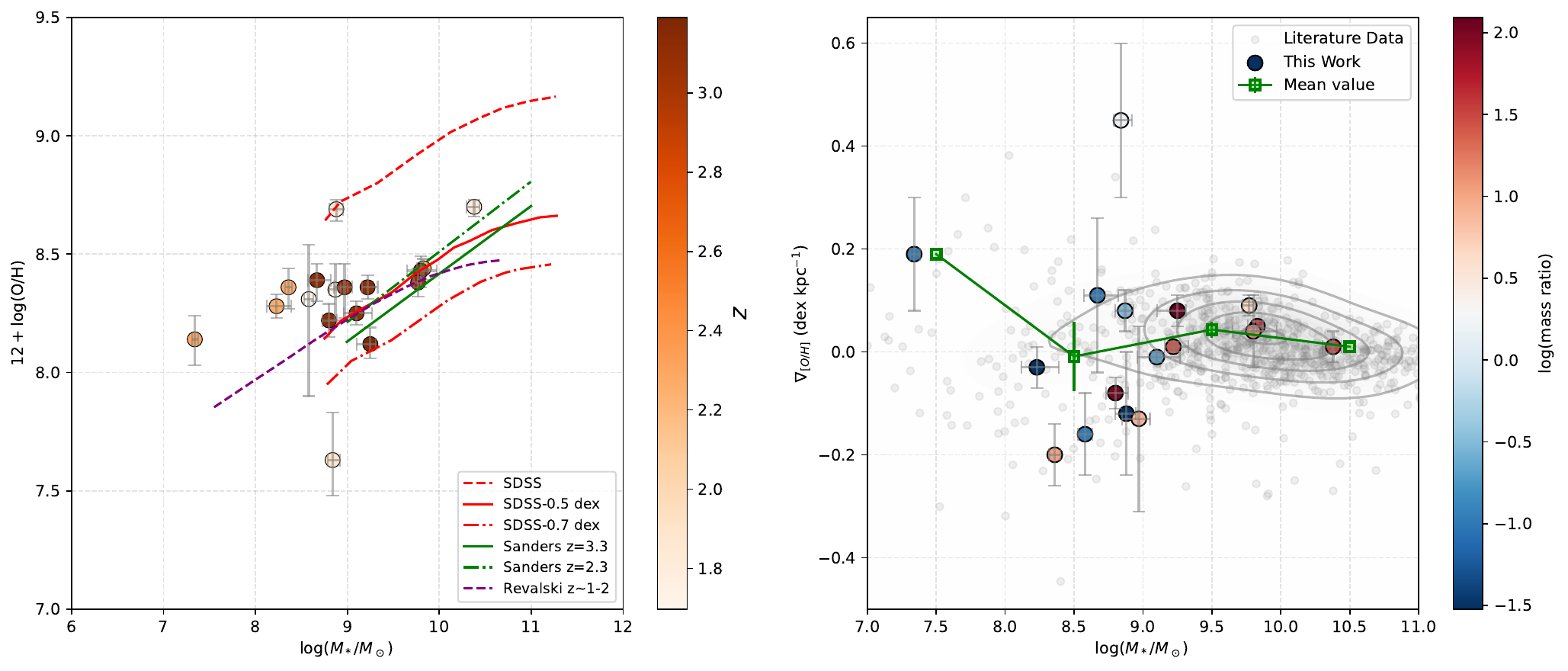}
    \caption{
Left: Stellar mass–metallicity relation (MZR) for galaxies in our sample, with points color-coded by redshift. Error bars reflect 1-$\sigma$ uncertainties in both stellar mass and gas-phase metallicity. Overplotted are empirical MZR relations from the literature, including SDSS-based calibrations and high-redshift measurements from \citet{Tremonti2004MZR,Li2023dwarfMZR,Sanders2021, 2024Revalski}.  
Right: Metallicity gradient ($\nabla$[O/H]) as a function of stellar mass. Our measurements are highlighted by big circles color-coded by mass ratio, with error bars reflecting 1-$\sigma$ uncertainties. The green squares indicate binned mean values with 1$\sigma$ errors. Gray background points represent literature data compiled from \citet{1992Vila-Costas,1994Zaritsky,2012Swinbank,2013Jones,2015Jones,2015Ho,2016Leethochawalit,2016Wuyts,2017Molina,2018Carton_z0.8-1,2018Forster,2019Patricio,2021Simons,2024Venturi,vanZee1998}. The background contours show a KDE-smoothed distribution for literature data at $0 < z < 3.2$. Both panels illustrate the dependence of chemical enrichment and radial gradients on stellar mass and pair properties.
}

    \label{fig:Zg_M_mr} 
\end{figure*}

\begin{figure*}
    \centering
    \includegraphics[width=1\linewidth]{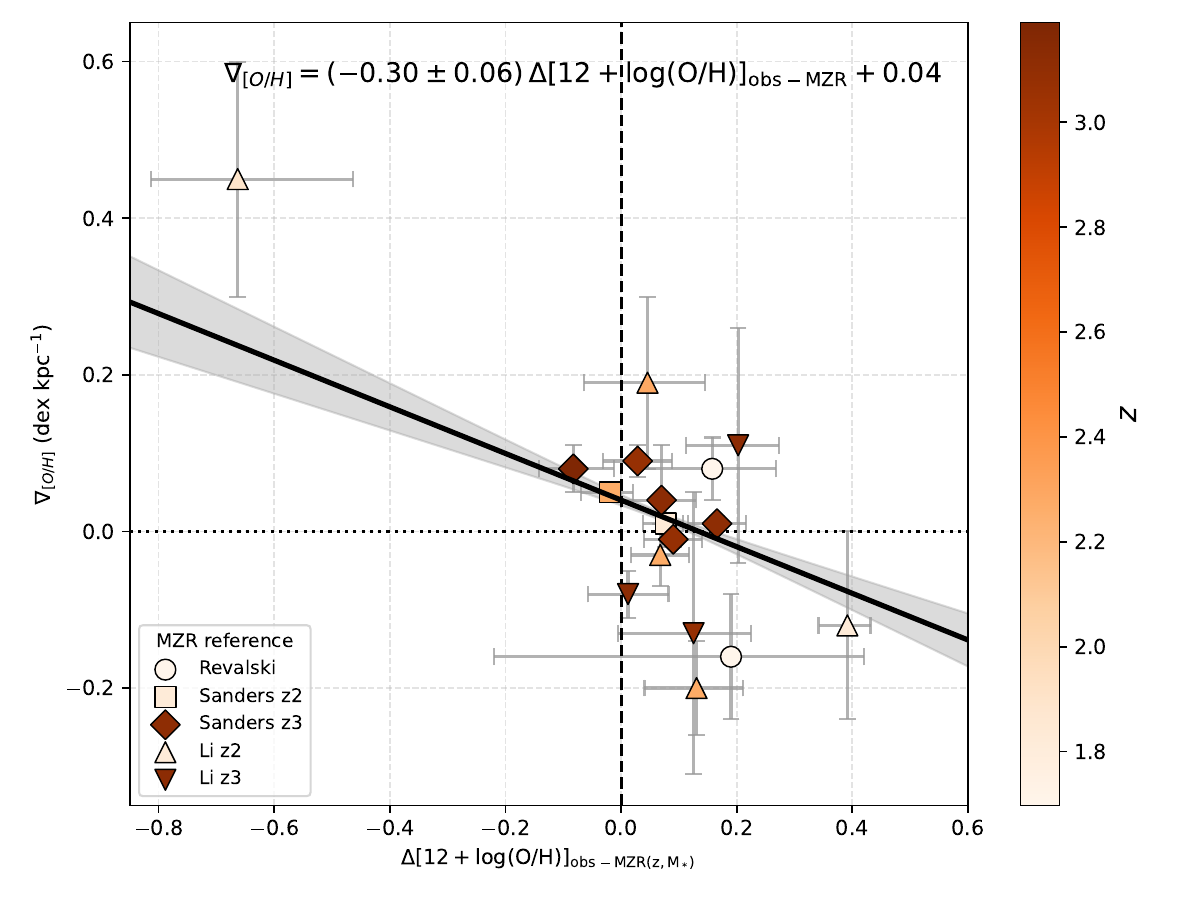}
\caption{
Metallicity gradient $\nabla_{\rm [O/H]}$ (dex\,kpc$^{-1}$) as a function of the metallicity offset relative to the redshift- and stellar-mass--matched mass--metallicity relation,
$\Delta[12+\log(\mathrm{O/H})]_{\rm obs-MZR}$.
For each galaxy, the reference MZR is selected according to its redshift and stellar mass, adopting (in order of priority) the relations from \citet{2024Revalski} for $0.8 \le z < 1.8$, \citet{Sanders2021} for $1.8 \le z < 4.0$, and \citet{Li2023dwarfMZR} where applicable.
Marker shapes identify the adopted MZR reference, as indicated by the legend.
The color scale denotes redshift.
The black solid line shows the weighted linear fit to the valid sample, accounting for uncertainties in the metallicity gradients, with the shaded region indicating the $1\sigma$ uncertainty of the fit.
Dashed and dotted lines mark $\Delta[12+\log(\mathrm{O/H})]=0$ and $\nabla_{\rm [O/H]}=0$, respectively.
}
    \label{fig:Zg_dz_z} 
\end{figure*}

\begin{figure*}
    \centering
    \includegraphics[width=1\linewidth]{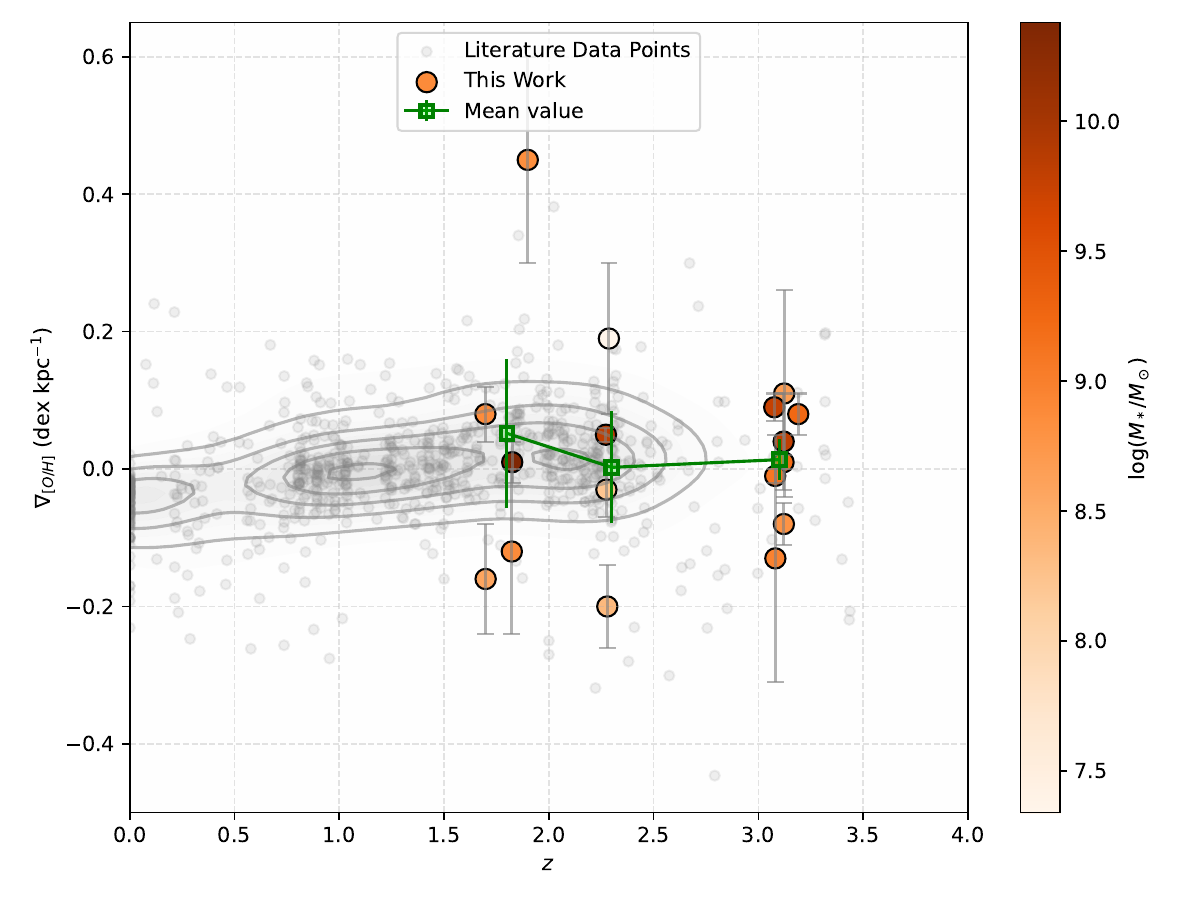}
    \caption{
Redshift evolution of \Zg for interacting galaxies.
The background grayscale points represent literature data from the compilation by the
same works in Figure \ref{fig:Zg_M_mr}, showing the general trend across cosmic time.
Orange circles denote individual galaxies from this work, color-coded by stellar mass ($\log(M_*/M_\odot)$).
Green squares represent the mean values of our sample in redshift bins, with error bars showing the standard deviation within each bin.
}

    \label{fig:Zg_z_M} 
\end{figure*}

\begin{figure*}
    \centering
    \includegraphics[width=1\linewidth]{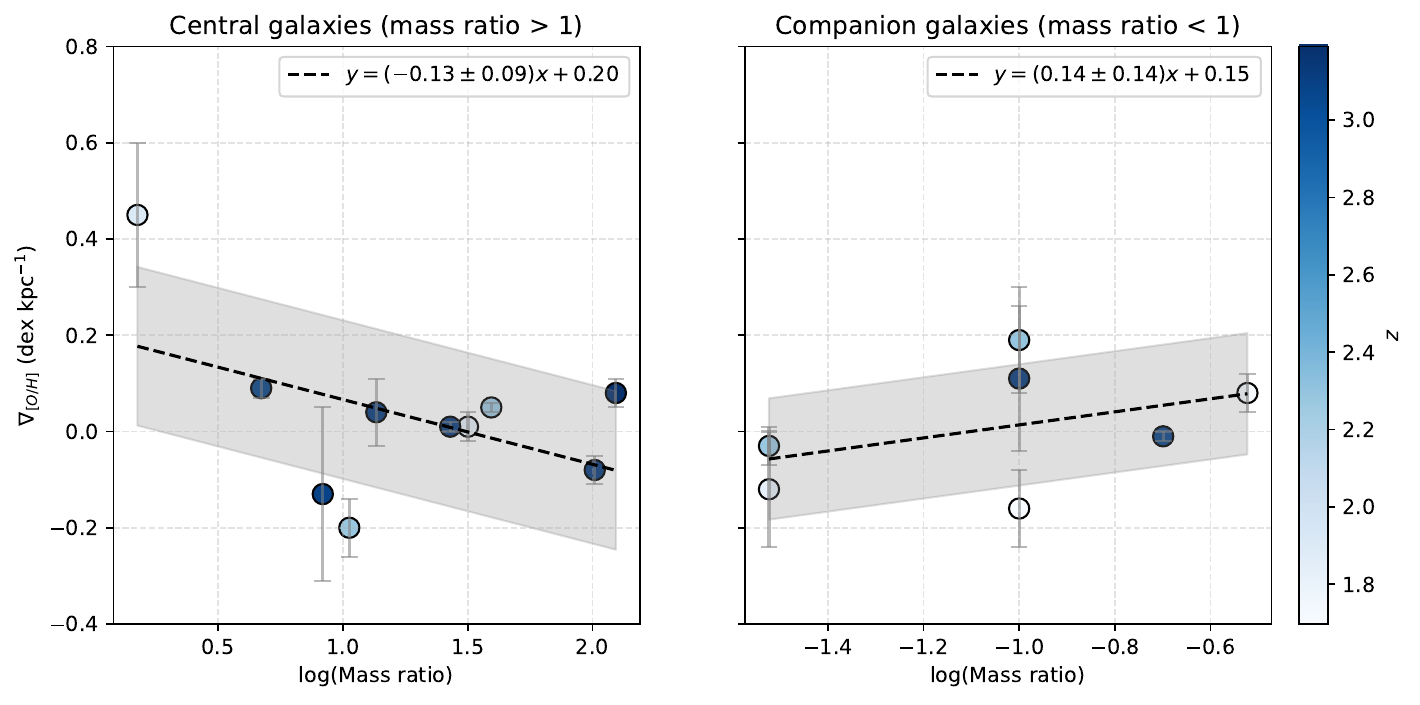}
    \caption{Metallicity gradients (\Zg, in dex~kpc$^{-1}$) as a function of stellar mass ratio ($M_*^{\rm self}/M_*^{\rm comp}$) for galaxies in interacting systems. 
The left and right panels show the central/primary (more massive) and companion/secondary (less massive) galaxies, respectively. 
Each point represents an individual galaxy, color-coded by redshift. 
Dashed lines indicate linear fits to the data in logarithmic mass-ratio space. 
The figure illustrates the weak, statistically non-significant trends for primary and secondary galaxies. 
}

    \label{fig:Zg_mr} 
\end{figure*}

We use the {\tt AstroPhot} \citep{Stone2023} package to model the galaxy light distributions in the JWST F115W, F150W, F277W, F444W, and HST F814W bands (see Section~\ref{sec:datsam}), fully accounting for PSF convolution. Each galaxy is modeled with a single Sérsic profile:
\begin{equation}
I(r) = I_e \exp\left\{ -b_n \left[ \left( \frac{r}{r_e} \right)^{1/n} - 1 \right] \right\}.
\end{equation}

Here, $r$ is the elliptical radius, $I_e$ is the surface brightness at
the effective radius $r_e$, $n$ is the Sérsic index, and $b_n$ is the
$n$-dependent normalization coefficient chosen such that $r_e$
encloses half of the total model light.

The fitting is performed simultaneously for all galaxies in each image and across all bands, yielding a self-consistent description of both structural parameters and total fluxes. Structural parameters are primarily constrained using  F277W  . This band samples  rest-frame optical emission   at $z\gtrsim2.6$ and rest-frame near-infrared emission at lower redshift, and in either case provides a less dust-sensitive  proxy for the stellar   distribution than the shorter-wavelength bands .

The Sérsic index $n$ characterizes the concentration of the light profile, with $n=1$ corresponding to an exponential disk and $n=4$ to a de Vaucouleurs profile. The position angle (PA) and axis ratio ($b/a$) describe the projected orientation and ellipticity. Both are fitted as part of the two-dimensional elliptical Sérsic model and enter through the definition of the elliptical radius $r$ in the above equation. In practice, the measured $b/a$ also reflects the relative contributions of structural components such as bulges and disks. The galaxy center is defined from the best-fit model.

From the best-fit models, we derive structural parameters ($r_e$, PA, $b/a$, and center) as well as total fluxes in all bands, enabling consistent multi-wavelength photometry for subsequent SED fitting. The $1\times$ and $2\times\,r_e$ contours are shown in Figure~\ref{fig:1}, and the structural parameters measured in the F277W band are listed in Table~\ref{tab:2}.

  Using $n=2$ as a descriptive division \citep{2010Fisher}, 15 of 17 galaxies are  disk-like and   two are more centrally concentrated. We use this result only to motivate a disk-based deprojection; a single  Sérsic index   is not a definitive dynamical classification, especially for disturbed  systems.

\subsection{SED Fitting} \label{subsec:sed}

We use the photometry obtained in the previous section to estimate the stellar masses of galaxies through spectral energy distribution (SED) fitting. This is performed using the Bayesian SED-fitting code \textsc{Bagpipes} version 1.0.1 \citep{2018Carnall}. The fitting procedure adopts the stellar population synthesis models of \citet{Bruzual2003} and applies the dust attenuation law of \citet{Calzetti2000}. Nebular continuum and emission lines are modeled using the photoionization code \textsc{Cloudy} \citep{2017FerlandRMxAA..53..385F}, following the methodology of \citet{2017Byler}. We assume a \citet{Chabrier2003} initial mass function (IMF) and employ double-power-law star formation histories. By this process, we derive stellar masses for all galaxies in our sample, as listed in Table \ref{tab:2}. We list all our metallicity available samples in Table \ref{tab:2}, which is subsample of Table \ref{tab:1} because of the SNR requirements for emission lines.

\subsection{Metallicity, dust extinction, and SFR}  \label{subsec:M}

We derive the gas-phase metallicity (\Z), nebular dust extinction ($A_V$), and intrinsic H$\beta$ flux ($f_{\mathrm{H}\beta}$) using the Bayesian forward-modeling approach presented in \citet{2017xwang,2019XWang,2020XWang,2022XWang_glass,2022XWng_mzr}. This method employs a likelihood function of the form:

\begin{equation}\label{eq:chi2}
    \mathcal{L} \propto \exp\left( -\frac{1}{2} \cdot \sum_i \frac{ \left( f_{\mathrm{EL}_i} - R_i \cdot f_{\mathrm{H}\beta} \right)^2 }{ \sigma_{\mathrm{EL}_i}^2 + f_{\mathrm{H}\beta}^2 \cdot \sigma_{R_i}^2 } \right)
\end{equation}

where $f_{\mathrm{EL}_i}$ and $\sigma_{\mathrm{EL}_i}$ denote the dereddened flux and associated uncertainty of each emission line (e.g., [O\,\textsc{ii}], H$\gamma$, H$\beta$, [O\,\textsc{iii}]), respectively. The fluxes are corrected for dust attenuation using the \citet{Calzetti2000} extinction law, with $A_V$ treated as a free parameter. The uncertainties $\sigma_{\mathrm{EL}_i}$ are obtained from the flux-calibrated error arrays \citep[see][]{Roberts2022}.

The quantity $R_i = f_{{\rm EL}_i} / f_{\mathrm{H}\beta}$ represents the expected flux ratio of each emission line relative to H$\beta$, based on the adopted metallicity calibration and intrinsic Balmer decrement, assumed to be ${\mathrm{H}\gamma}/{\mathrm{H}\beta} = 0.47$. The term $\sigma_{R_i}$ accounts for the intrinsic scatter in $R_i$, including uncertainties from the calibration models.

To estimate gas-phase metallicities, we adopt the empirical strong-line diagnostics developed by \citet{2018Bian}, which are calibrated against local analogs of high-redshift galaxies selected via their positions on the BPT diagram \citep{Baldwin1981}. In our analysis, the dust extinction $A_V$ is constrained not only by Balmer line ratios (e.g., H$\gamma$/H$\beta$, H$\alpha$/H$\beta$) but also by its influence on the metallicity diagnostics. For example, the empirical correlation between $R_{23}$ and $O_{32}$ \citep[e.g.,][]{2015Shapley}, where $R_{23} = ([\mathrm{O\,II}] + [\mathrm{O\,III}])/\mathrm{H}\beta$ and $O_{32} = [\mathrm{O\,III}]/[\mathrm{O\,II}]$, as well as intrinsic line ratios such as [O\,\textsc{iii}]/[Ne\,\textsc{iii}] \citep{2015Jones}, provides additional constraints on $A_V$. We propagate the uncertainty in $A_V$ into the final metallicity uncertainties.

 \textcolor{black}{The available line set does not include a statistically useful auroral line such as [O\,\textsc{iii}]~$\lambda4363$, so a direct-$T_e$ measurement is not possible even for a representative subset. Nevertheless, JWST direct-$T_e$ measurements now provide high-redshift strong-line calibrations over a redshift range overlapping our sample \citep{Scholte2025EXCELS,Sanders2026AURORA}. We therefore performed a homogeneous calibration-sensitivity test for all 17 galaxies using the \citet{Sanders2026AURORA} $O2=[\mathrm{O\,II}]\lambda\lambda3727,3730/\mathrm{H}\beta$ and $O3=[\mathrm{O\,III}]\lambda5008/\mathrm{H}\beta$ relations. The two relations were fitted jointly in flux space together with the intrinsic $\mathrm{H}\gamma/\mathrm{H}\beta=0.47$ ratio. We propagated the uncertainty of each line and the intrinsic scatter of the calibration. We did not add $R_{23}$ or $O_{32}$ as independent constraints in this test because they are constructed from the same line fluxes and would otherwise double-count correlated information.}

\textcolor{black}{The Sanders et al. calibration lowers the inferred metallicity by a median of approximately 0.15~dex at the galaxy level relative to \citet{2018Bian}. The radial gradients are less strongly affected: for the 17 galaxies with at least three usable spatial bins, the median Sanders-minus-Bian gradient change is $-0.012$~dex~kpc$^{-1}$ and the median absolute change is 0.019~dex~kpc$^{-1}$. The number of gradients detected at $\geq2\sigma$ changes from five to four. We retain the Bian et al. values as the primary measurements for consistency with our earlier NIRISS analysis, but use the Sanders et al. reanalysis to identify calibration-sensitive conclusions.}

 We apply the method described above to calculate the metallicity, $A_V$, and the dereddened $f_{\mathrm{H}\beta}$ for each galaxy, as listed in Table \ref{tab:2}. Since star formation is closely linked to metal enrichment, we also estimate the star formation rate (SFR) using the dust-corrected $\mathrm{H}\beta$ flux, following the calibration of \citet{Kennicutt1998} and adopting the \citet{Chabrier2003} IMF:
\begin{equation}
    \mathrm{SFR}\ [M_\odot\ \mathrm{yr}^{-1}] = 1.6 \times 10^{-42}\ L_{\mathrm{H}\beta}\ [\mathrm{erg}\ \mathrm{s}^{-1}].
\end{equation}
 We use H$\beta$ because it measures the ionizing output of massive stars and therefore probes very recent star formation (approximately the last $\lesssim10$ Myr), on a timescale directly relevant to an interaction-triggered burst. In contrast, the SFR implicit in the \textsc{Bagpipes} fit is averaged over the adopted star-formation history and is typically sensitive to a longer, prior-dependent interval of roughly 50--100 Myr. We therefore adopt H$\beta$-based SFRs in our analysis; this timescale distinction prevents us from interpreting a displacement as a sustained enhancement without a matched control sample.

\subsection{Metallicity radial gradient}  \label{subsec:MG}
The NIRISS slitless spectroscopy provides spatially resolved   emission-line maps. Because the line flux is distributed over multiple spatial pixels, each pixel receives only a small fraction  of the total   signal. The resulting per-pixel SNR is therefore much lower than that of the integrated spectrum .

To improve the SNR, we apply Voronoi binning to the data. Each bin is required to satisfy $\mathrm{SNR}_{\mathrm{[O\,III]}} > 5$, ensuring sufficient signal strength for reliable analysis. This method enhances the detectability of faint emission in the galaxy outskirts and allows for a more complete mapping of the ionized gas, albeit at the sacrifice of some spatial resolution. 

 \textcolor{black}{We characterize the physical size of each Voronoi bin using the diameter of a circle with the same projected area, $D_{\rm eq}=2\sqrt{A_{\rm bin}/\pi}$, where $A_{\rm bin}$ is calculated from the number of $0.''06$ pixels and the angular-diameter distance at the redshift of each galaxy. Across the 658 spatial bins in the quantitative sample, the mean equivalent diameter is 0.82~kpc and the median is 0.59~kpc, with a 16th--84th percentile range of 0.53--1.06~kpc. These values describe the spatial sampling of the Voronoi maps rather than the number of fully independent resolution elements, which is limited by the NIRISS PSF.}

 We apply our Bayesian inference method to each Voronoi bin to construct spatially resolved gas-phase metallicity (\Z) maps for the target galaxies, as shown in Figure~\ref{fig:2}. Based on these \Z\ maps and the derived morphological parameters, we measure the radial metallicity gradients.
Because the \Z\ maps are projected along the line of sight, we   deproject  the spatial distribution   using the position angle and observed axis ratio from the morphological fit. We assume a thin disk ($q_0=0$); adopting a finite intrinsic thickness would instead reduce the inferred inclination through $\cos^2 i=[(b/a)^2-q_0^2]/(1-q_0^2)$. For disturbed galaxies, neither prescription fully describes the three-dimensional structure, so the deprojected radii---especially for apparently elongated objects---should be regarded as model dependent .

We then plot the metallicity of each Voronoi bin as a function of the deprojected galactocentric distance, which is determined from the morphological fitting, as shown in the right panels of Figure~\ref{fig:2}. Each blue data point, together with its error bar, represents an individual Voronoi bin. We perform a weighted least-squares fit to derive the metallicity gradient, together with its 1$\sigma$ uncertainty, which is reported in the text box of each panel. Hereafter, we use \Zg\ to denote the slope of \Z\ with respect to galactocentric radius. Although $\nabla$ formally represents the spatial gradient operator, in this work we consider only the radial component. The best-fit gradient is shown as a red dashed line, while the shaded red region indicates the 95\% confidence interval. For reference, the locations of 1$R_e$ and 2$R_e$ are marked to illustrate the radial range over which the gradient is measured.

 We tested the robustness of the inferred metallicity gradients by applying additional S/N requirements to the emission lines used in the metallicity inference. Our baseline measurements use Voronoi bins selected at ${\rm S/N}([{\rm O\,III}])\geq5$, without imposing an additional bin-level S/N threshold on [O~II] or H$\beta$. Requiring ${\rm S/N}([{\rm O\,II}])\geq2$ leaves 16 galaxies with at least three spatial bins; all 16 gradients are consistent with their baseline values within $1\sigma$, with a median absolute change of $0.008~{\rm dex\,kpc^{-1}}$. With the more restrictive requirement ${\rm S/N}([{\rm O\,II}])\geq3$, 15 galaxies remain measurable, of which 14 are consistent with the baseline measurements within $1\sigma$, and the median absolute change is $0.013~{\rm dex\,kpc^{-1}}$. A few best-fit gradients change sign, but most of these measurements are consistent with zero; only Galaxy~783 changes by more than $2\sigma$ under the ${\rm S/N}([{\rm O\,II}])\geq3$ selection. More stringent cuts on H$\beta$ or [O~III] preferentially remove low-surface-brightness bins and substantially reduce the radial coverage, and can therefore alter individual gradients through spatial-selection effects. We consequently adopt the [O~II]-based tests as the primary robustness check. These tests show that the sample-level gradient distribution is not systematically driven by low-S/N [O~II] measurements, while individual gradients may remain sensitive to the adopted spatial-bin selection.

Figure~\ref{fig:snr_robustness} in the Appendix summarizes these selection tests for the 17-galaxy quantitative sample. The existing bin-level metallicity posteriors are held fixed, and the radial gradients are refitted after each additional S/N cut.

We report gradients in both dex kpc$^{-1}$ and dex $R_e^{-1}$ in Table~\ref{tab:2}; the latter is obtained by multiplying the physical gradient by $R_e$ and facilitates comparisons across the broad size range of the sample.

\subsection{Spatial-resolution and beam-smearing limitations}
\begin{figure*}[htbp]
\centering
\includegraphics[width=.95\textwidth]{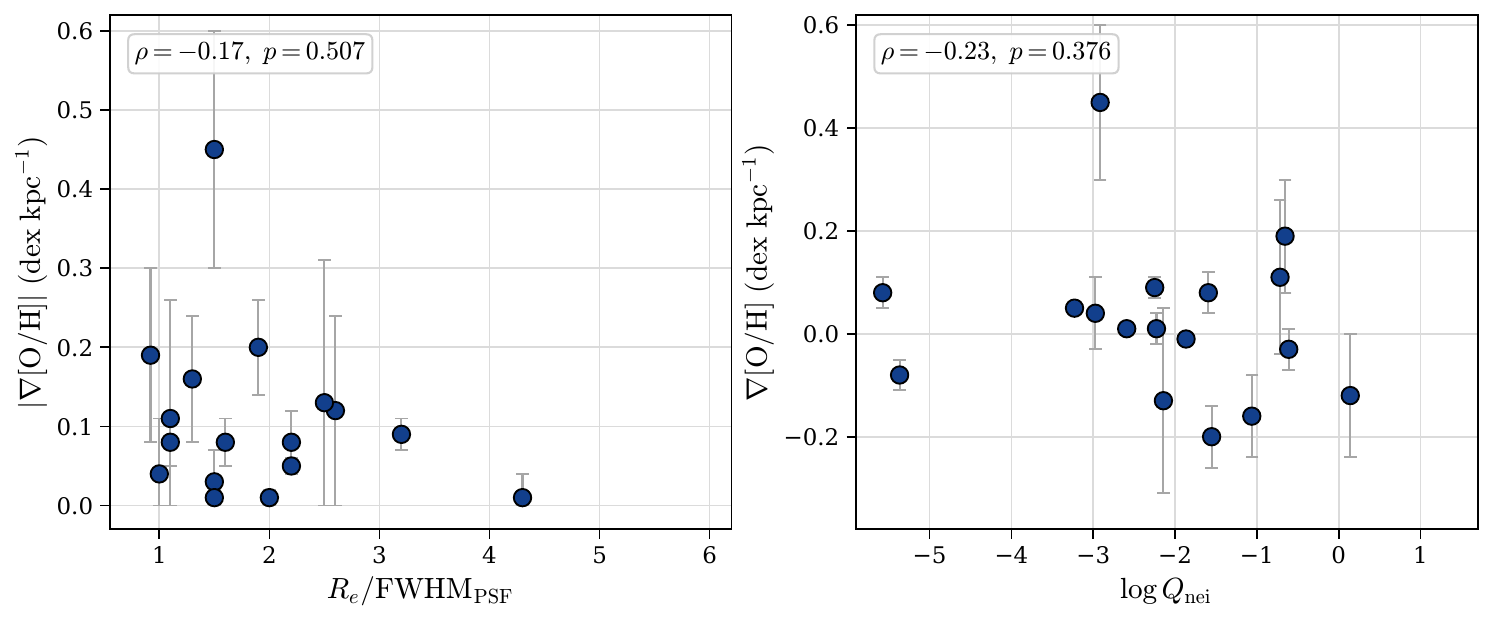}
\caption{Resolution and interaction-strength diagnostics for the 17-galaxy quantitative sample. Left: absolute metallicity-gradient magnitude, $|\nabla_{\rm [O/H]}|$, versus effective radius normalized by a representative NIRISS PSF FWHM of $0.10\arcsec$. The lower error-bar bounds are truncated at zero because the plotted quantity is an absolute magnitude. Right: signed metallicity gradient versus the neighbor tidal-strength parameter. Neither relation is statistically significant.}
\label{fig:resolution_tidal}
\end{figure*}

The wavelength-dependent NIRISS point-spread function (PSF) is comparable to the sizes of several targets. As a transparent diagnostic, we adopt a representative FWHM of $0.10\arcsec$. Because beam smearing drives intrinsic gradients of either sign toward zero, we plot the absolute magnitude of the measured metallicity gradient, $|\nabla_{\rm [O/H]}|$, against $R_e/{\rm FWHM}_{\rm PSF}$ in Figure~\ref{fig:resolution_tidal}. The most compact objects retained in the quantitative sample (1308 and 2252) have $R_e/{\rm FWHM}_{\rm PSF}\lesssim1$. Beam smearing mixes light from different intrinsic radii and generally suppresses the magnitude of a resolved gradient \citep{Yuan2011,Wuyts2016b,2018Carton_z0.8-1,Acharyya2020}; we have not forward-modeled or corrected this effect. The gradient magnitudes of these compact systems should therefore be interpreted as lower limits. For the 17-galaxy quantitative sample, a Spearman rank test gives $\rho=-0.17$ and $p=0.507$, providing no evidence for a monotonic correlation between $|\nabla_{\rm [O/H]}|$ and $R_e/{\rm FWHM}_{\rm PSF}$. Given the small sample, the uncertainties in the measured gradients, and our use of a representative rather than wavelength-specific PSF, this null result does not demonstrate that beam smearing is negligible.

 \subsection{Major Pair and Minor Companions}
\label{subsec:oii_sii}

In an interacting system, the stellar mass ratio of the galaxies plays a crucial role in determining the nature and outcome of their interaction. Major mergers, with mass ratios close to unity, are known to induce strong tidal features, enhanced central gas inflows, and intense star formation, while minor mergers typically lead to weaker morphological and kinematic disturbances \citep{1972Toomre, 1992Barnes, Cox2008massratio_starburst, 2009Hopkins, 2010Lotz_mr}.

Motivated by these results, we examine whether the stellar mass ratio is related to the observed metallicity gradients in our sample. We define the stellar mass ratio as
\begin{equation}
    \frac{M_{*}^{\mathrm{self}}}{M_{*}^{\mathrm{comp}}},
\end{equation}
where values greater (less) than unity indicate that the galaxy is the more (less) massive member of the interacting pair. For each pair, Table~\ref{tab:2} also reports a single pair mass ratio, $\mu=M_{*,\mathrm{secondary}}/M_{*,\mathrm{primary}}\leq1$, together with the ID of the primary galaxy. In contrast, the galaxy-specific ratio $M_{*}^{\mathrm{self}}/M_{*}^{\mathrm{comp}}$ is used for the color coding in Figure~\ref{fig:Zg_M_mr} and for the correlation analysis below. Following the commonly adopted convention, systems with
\begin{equation}
    \frac{1}{4} \leq \frac{M_{*}^{\mathrm{self}}}{M_{*}^{\mathrm{comp}}} \leq 4
\end{equation}
are classified as major pairs, while those outside this range are classified as minor pairs.

We find that only 2 out of 17 galaxies in our sample are involved in major pairs, whereas the remaining 15 galaxies are in minor interactions. Notably, Galaxy~1785, which exhibits the most positive metallicity gradient in our sample ($\sim 0.45~\mathrm{dex~kpc^{-1}}$), belongs to a major pair (1765--1785) with a stellar mass ratio of $\sim 1.5$.

 Quantitative tidal criteria based on the relative masses, separation, and size of interacting galaxies provide a more objective assessment of interaction strength than morphology alone \citep{HenriksenByrd1996,2025Watson}. Similar approaches have been used to distinguish tidal perturbations from other environmental processes \citep{Merluzzi2016,Lassen2026}. Because mass ratio alone omits separation and galaxy size, we follow commonly used estimates of the tidal influence of neighboring galaxies \citep{Athanassoula1984,Verley2007,ArgudoFernandez2014} and calculate the neighbor tidal-strength parameter
\begin{equation}
\log Q_{\rm nei}=3\log_{10}\left(\frac{R_{e,{\rm targ}}}{d_{\rm proj}}\right)-\log_{10}\left(\frac{M_{\star,{\rm targ}}}{M_{\star,{\rm nei}}}\right).
\end{equation}
Here $R_e$ and $d_{\rm proj}$ are in the same physical units. Larger $Q_{\rm nei}$ denotes a more massive and/or closer perturber relative to the target size. The right panel of Figure~\ref{fig:resolution_tidal} shows no significant monotonic relation with the signed metallicity gradient ($\rho=\rhoTidal$, $p=\pTidal$). This remains a pairwise estimate: the available spectroscopic catalog is selected for strong emission lines and is not demonstrably complete to a common stellar-mass limit around every target. Summing over all cataloged neighbors would therefore produce a selection-dependent ``total'' tidal field, so we do not interpret it as an unbiased environmental metric.

 \section{Results and Discussion}     \label{sec:rslt}

\subsection{Star Formation Rate in Interacting Galaxies} 

  \textcolor{black}{The star-formation response to galaxy interactions is not uniform and depends on redshift, separation, and merger stage. At $0.5<z<3.0$, \citet{Shah2022} found modest SFR enhancements in close spectroscopic pairs and stronger enhancement in visually identified coalesced systems, although both were weaker than those measured for comparable local interactions. By contrast, \citet{Silva2018} found no significant difference between the star-forming properties of merging and non-merging galaxies at $0.3<z<2.5$, while \citet{Pearson2019} measured only a small average enhancement (a factor of approximately 1.2) across a large sample extending to $z=4$. Moreover, the interaction sequence need not end in sustained enhancement: \citet{Ellison2022} found post-starburst signatures to be 30--60 times more frequent in post-merger remnants than in matched controls, with no corresponding excess in close pairs, indicating that rapid quenching can emerge after coalescence. These results emphasize that enhanced, unchanged, and subsequently quenched star formation can represent different systems or phases of the merger sequence.}

 \textcolor{black}{Figure~\ref{fig:SFMS} compares the H$\beta$-based SFRs of our metallicity sample with the redshift-dependent star-forming main-sequence relations of \citet{Speagle2014}. The sample points are color-coded by their spectroscopic redshifts to facilitate comparison with the appropriate redshift-dependent relation. Galaxy~2252 appears to lie above the corresponding relation, whereas Galaxy~783 lies below it. The remaining galaxies are broadly consistent with the main sequence within their uncertainties. Therefore, Figure~\ref{fig:SFMS} alone does not provide evidence for a systematic enhancement of star formation in the pair-selected sample.}

\textcolor{black}{Establishing an interaction-induced SFR enhancement would require a control sample of isolated galaxies matched in redshift, stellar mass, emission-line sensitivity, and data quality, and analyzed using the same methodology. Such a statistically meaningful control sample cannot be constructed from the dataset analyzed here. We therefore treat the comparison with the star-forming main sequence as descriptive and do not infer a significant merger-versus-non-merger difference in SFR.}

 \subsection{Metallicity Gradient vs. Stellar Mass} 

Previous studies have found that merging galaxies can deviate from the standard mass--metallicity relation (MZR) \citep[e.g.,][]{Tremonti2004MZR}. To investigate whether similar offsets occur in our sample, we examine the MZR for our candidate interacting systems in the left panel of Figure~\ref{fig:Zg_M_mr}. The integrated gas-phase metallicity is derived from emission-line fluxes measured from the one-dimensional spectra extracted with Grizli. The color-coded points with error bars represent our galaxy sample, while we overplot empirical MZRs from the literature: the $z\sim0$ SDSS MZR \citep{Tremonti2004MZR}, the $z\sim 1-2$ MZR from \citet{2024Revalski} (which is also similar to the correlation found by \citet{Henry2021}), the $z\sim2$ MZR from \citet{Sanders2021}, and the $z\sim3$ MZR from \citet{Li2023dwarfMZR}.

We find that many of our galaxies lie close to the expected MZRs. In particular, the galaxies at $z\sim2.2$ agree well with \citet{Li2023dwarfMZR} at $\log(M_*/M_\odot)<9$ and with \citet{Sanders2021} at $\log(M_*/M_\odot)>9$, while one galaxy at $z\sim3$ lies above the \citet{Sanders2021} relation. We note that merging galaxies were not explicitly excluded from the samples of either \citet{Li2023dwarfMZR} or \citet{Sanders2021}, which may introduce additional scatter in their MZR relations. Some galaxies, especially those at $z\sim3$, deviate from the corresponding MZR and appear as metallicity outliers. Gas transfer, stripping, accretion, or radial inflow associated with an interaction could contribute to such offsets by redistributing gas and metals within or between galaxies \citep{2010Rupke,2022XWang_glass,2024Venturi}. However, the present sample has no matched control and does not uniquely identify any of these mechanisms.

Accordingly, we regard the MZR offsets as descriptive evidence of chemical diversity within the pair-selected sample, rather than as proof that interactions caused the observed scatter. Larger matched samples and resolved gas kinematics are required to distinguish interaction-driven redistribution from intrinsic scatter, measurement systematics, and selection effects.

  We next  investigate the relationship between gas-phase metallicity   gradient  (\Zg) and stellar mass (\mstar)  in the right panel of Figure~\ref{fig:Zg_M_mr}.   To match the epoch studied here, the displayed comparison is restricted to $0<z<3.2$ \citep{1992Vila-Costas,1994Zaritsky,2012Swinbank,2013Jones,2015Jones,2015Ho,2016Leethochawalit,2016Wuyts,2017Molina,2018Carton_z0.8-1,2018Forster,2019Patricio,2021Simons,2024Venturi,vanZee1998}. The literature compilation is heterogeneous: metallicity calibrations, PSF sizes, gradient-fitting ranges, IMF assumptions, and merger classifications are not uniform. Where conversions are published, stellar masses are placed on a Chabrier IMF; entries without sufficient information are left as reported. Consequently, the background distribution provides context rather than a matched non-interacting control, and small offsets should not be interpreted physically.

We divide our sample into four stellar mass bins and plot the mean \Zg values in each bin using green error bars, which indicate the standard deviation ($1\sigma$).   The mean gradients  tend to be flat or mildly positive  . Although  mixing or inflow   in interacting systems could produce such  gradients \citep{2024Venturi}  , the absence of a matched control sample prevents us from attributing this behavior specifically to interactions. The sample appears to have  a larger scatter   than the literature compilation, but bootstrap resampling gives a  95\% confidence interval of  $[-0.002,0.126]$ for  the scatter difference , consistent with no significant difference  . The heterogeneous literature compilation therefore provides context only and cannot substitute for an isolated control sample. \textcolor{black}{The direct-$T_e$-anchored calibration test described in Section~3.3 supports this population-level interpretation: most gradients remain statistically consistent between the Bian and Sanders calibrations, and there is no systematic reversal of the sample distribution. Individual objects are less robust. In particular, for Galaxy~913 the uniformly refitted gradient changes from $+0.032\pm0.006$ to $+0.010\pm0.008$~dex~kpc$^{-1}$ and is no longer significant. Galaxies~1484, 2251, and 2303 retain significant gradients under both calibrations. Galaxy~2649 is sensitive to spatial bins close to the upper validity boundary of the Sanders et al. calibration; we therefore do not regard its calibration-dependent significance as robust.}

Related observational and theoretical work likewise shows that interactions can redistribute gas and metals and thereby alter radial abundance profiles \citep{2010Rupke,2022XWang_glass,2024Venturi}.

To further investigate whether the internal metallicity structure of interacting galaxies is connected to their global chemical enrichment, Figure~\ref{fig:Zg_dz_z} shows the gas-phase metallicity gradient as a function of the metallicity offset relative to the redshift- and stellar-mass--matched mass--metallicity relation, $\Delta[12+\log(\mathrm{O/H})]_{\rm obs-MZR(z,\,M_{*})}$.
Overall, galaxies with negative metallicity offsets tend to exhibit flatter or even positive metallicity gradients, whereas galaxies with positive metallicity offsets show a broader distribution of gradient values, including both flat and strongly negative gradients. A weighted linear fit to the 17-galaxy quantitative sample yields a formal relation of
$\nabla_{\rm [O/H]} = (-0.30\pm0.06) \, \Delta[12+\log(\mathrm{O/H})]_{\rm obs-MZR(z,\,M_{*})} + 0.04,$ where the quoted uncertainty accounts for the metallicity-gradient measurement errors only. The dispersion about the fit exceeds that expected from these errors ($\chi^2_\nu=3.98$), indicating additional scatter. Non-parametric tests provide only marginal evidence for an anti-correlation: Spearman's $\rho=-0.45$ ($p=0.071$) and Kendall's $\tau=-0.34$ ($p=0.058$). The result is also sensitive to individual objects and is therefore treated as tentative rather than as a robust $5\sigma$ detection. A similar relation has been explored by \citet{2025Li_Tie} for dwarf galaxies, where a correlation ($-0.14 \pm 0.05$) between metallicity gradients and offsets from the MZR in inner region was reported. They interpret this trend as a consequence of gas inflow and feedback-driven metal redistribution, which suppress central metal enrichment and lead to flatter gradients.

For our sample of star-forming galaxy pairs, we find that galaxies lying below the MZR tend to exhibit flatter or even positive metallicity gradients. Although the correlation is not statistically significant given the current sample size, this trend is qualitatively consistent with a scenario in which galaxy interactions enhance the inflow of metal-poor gas and promote metal mixing, preferentially reducing the central metallicity and thereby flattening the overall metallicity gradient. Such dilution-dominated phases and related departures from the mass--metallicity relation have been widely reported in local interacting and chemically unusual systems (e.g., \citet{Ellison2008,Peeples2009,2010Kewley,2018Bustamante}).

Our sample contains more galaxies on the metal-enhanced side of the adopted reference MZR than below it. Several non-unique effects could contribute to this distribution, including intrinsic scatter or systematic differences among metallicity calibrations, selection effects, enriched inflow, metal retention, or enhanced star formation and recycling. High-redshift galaxies generally have high gas fractions and short depletion timescales (e.g., \citet{2010Tacconi,2010Daddi,2010Genzel}), so interaction-associated inflow and star formation are also plausible contributors, but their importance cannot be established from these data alone.

The tentative anti-correlation between metallicity gradients and MZR offsets may therefore be compatible with a mixture of dilution, mixing, enrichment, and feedback-driven redistribution. Galaxies below the MZR could be experiencing dilution, whereas positive offsets could be associated with enrichment or with systematic and selection effects; these locations do not establish a unique evolutionary sequence. Such possibilities are qualitatively consistent with the diverse chemical behavior reported in galaxy pairs by \citet{2022A&A...660A.145O}.

Finally, we note that observational selection effects may also influence the observed distribution of metallicity offsets in our sample. Metal-poor systems at high redshift may exhibit fainter nebular emission and lower signal-to-noise ratios in key diagnostic lines, potentially making low-metallicity interacting galaxies more difficult to identify and characterize. As a result, the observed prevalence of metal-enhanced systems may be partially affected by selection biases. Larger and more uniformly selected samples spanning a wider range of galaxy properties and merger stages will therefore be required to robustly constrain the relative importance of dilution and enrichment processes across cosmic time.

\subsection{Metallicity Gradient vs. Redshift}
A   growing body of ground-based and JWST work has measured metallicity gradients beyond the local Universe, including studies that distinguish disks from interacting or kinematically disturbed systems \citep{2012Queyrel,Troncoso2014,2012Swinbank,2018Carton_z0.8-1,2018Forster,Gillman2020,2021Simons,2024Venturi,2025Ju,Lee2026,Stanghellini2026,li202513billionyearviewgalaxygrowth}. We therefore do not claim that interaction-aware measurements at high redshift are unprecedented. Instead, our contribution is a uniformly selected NIRISS pair sample concentrated at cosmic noon.

In this work, we compare our pair-selected sample of candidate interacting systems with results from the literature as a function of redshift, as shown in Figure~\ref{fig:Zg_z_M}. The orange data points with error bars represent our pair-selected sample, while the gray points correspond to literature data, consistent with those shown in Figure~\ref{fig:Zg_M_mr}.

We tentatively find that the scatter in \Zg\ among pair systems appears to be larger at $M_{*} < 10^9\,M_{\odot}$, although this result is subject to large uncertainties and small-number statistics.
Low-mass galaxies have shallower gravitational potentials and could therefore respond more strongly to gas inflows, outflows, or transfer between companions. These processes offer possible explanations for the apparent scatter, but the current sample is too small to establish either a mass dependence or an interaction-driven origin.

To better illustrate the trend, we compute the mean \Zg\ in three redshift bins at $z\sim 1.8, 2.3, 3.1$ and plot them as green squares. The mean gradients in all three redshift bins are consistent with zero within their uncertainties.
  The bin means are consistent with flat average metallicity gradients, which could arise from efficient metal mixing, although other explanations cannot be excluded. Individual cosmic-noon galaxies can nevertheless have negative gradients, so these measurements do not imply that negative disk gradients were absent. Merger-driven mixing operates at all epochs, while its incidence and persistence may increase at cosmic noon because gas fractions and merger rates are higher; the present sample cannot separate that redshift dependence from its explicit pair selection .

\subsection{Metallicity Gradient vs. Mass Ratio}
We further test the correlation between metallicity gradient and mass ratio using non-parametric methods. 
We plot the relationship between \Zg and log(mass ratio) for   primary  galaxies (mass ratio $>1$) and   secondary  galaxies (mass ratio $<1$), as shown in Figure~\ref{fig:Zg_mr}.
The   fitted slopes are $-0.13\pm0.09$ for primaries and $0.14\pm0.14$ for secondaries. Taken together, as a pair approaches equal mass, the gradients of both members tend toward flatter or positive values. These trends are weak and should be treated as descriptive rather than detections.

Simulations suggest that gravitational interactions, gas transfer, and star-formation feedback in major mergers may produce flat or positive metallicity gradients \citep{Brian2025}. These processes could potentially contribute to the positive gradients observed in the two galaxies associated with major pairs in our quantitative sample. In minor interactions, the lower-mass companion may be more susceptible to tidal stripping, whereas the primary may experience weaker perturbations \citep{2007Di_Matteo}. However, gas inflow, stripping, and metal redistribution depend strongly on the interaction geometry and merger stage. Given the large scatter and the absence of statistically significant correlations in our sample, these possible interpretations remain tentative.

The fitted mass-ratio trends have large uncertainties. For the primary galaxies, Kendall's $\tau=-0.09$ ($p=0.792$) and Spearman's $\rho=-0.16$ ($p=0.655$). For the secondary galaxies, Kendall's $\tau=0.29$ ($p=0.544$) and Spearman's $\rho=0.21$ ($p=0.662$). The large $p$ values show that neither trend is statistically significant.
Further studies of larger sample size may help us better understand the metallicity distribution in merging systems.

\subsection{Selected Systems in the Sample}

The galaxies in our sample exhibit a wide variety of colors, sizes, and morphologies. To better understand the potential physical mechanisms at play, we examine them individually.

 To visualize directional variations in Figure~\ref{fig:2}, we calculate a deprojected azimuthal angle for the centroid of each Voronoi bin. Relative to the S\'ersic-model center, the centroid offsets are rotated into coordinates $(x',y')$ aligned with the photometric major and minor axes using the fitted position angle. Adopting the fitted projected axis ratio $q=b/a$, we define the deprojected radius and folded azimuth as $r_{\rm dep}=[x'^2+(y'/q)^2]^{1/2}$ and $\phi=\tan^{-1}[(|y'|/q)/|x'|]$, respectively. Thus, $\phi=0^\circ$ identifies the major-axis direction and $\phi=90^\circ$ the minor-axis direction. This calculation assumes that the projected ellipticity traces an intrinsically circular thin disk; the colors should therefore be interpreted as a geometric guide rather than a precise three-dimensional direction.

 One of the first features we note is the inhomogeneous metallicity distribution among galaxies in the pair-selected sample. In Figure~\ref{fig:2}, Galaxy 913 appears to show low-metallicity gas in its outskirts along the major axis, approximately aligned with the line connecting the centers of the two galaxies. Although the spectral resolution limits our ability to analyze the gas kinematics, the spatial variation in metallicity may indicate inflowing gas triggered by the interaction.

Along the short axis of Galaxy 913, we observe metal-rich outskirts and an extended structure, which could be associated with outflows perpendicular to the disk. Taken together, the observed metallicity gradient (\Zg) may result from a combination of co-planar inflows (producing a negative \Zg) and centrally driven outflows (producing a positive \Zg). These two opposing trends are visible in the top-left panel of Figure~\ref{fig:2}. 

The companion galaxy (Galaxy 912) shows relatively  metal-rich gas on the side facing Galaxy   913. Several scenarios remain possible: redistribution of  pre-enriched   material, inflow of already enriched gas, or selective removal/outflow of enriched central gas. The present data lack the resolved kinematics required to distinguish among them .

For the other systems, due to the diversity in galaxy sizes, mass ratios, projected separations, and relative velocities, it is difficult to identify a single dominant mechanism driving the observed metallicity gradients. This highlights the importance of considering the sample as a whole rather than focusing solely on individual systems.

In addition to the individual galaxies discussed above, we examined the full set of spatially resolved metallicity measurements   ( Appendix~\ref{fig:appendix}).   Among the five pairs for which gradients are measurable for both members, four have opposite best-fit signs. Several gradients are individually consistent with zero, however, so this count is descriptive rather than evidence for a population-level effect .

\raggedbottom
\section{Conclusion} \label{sec:conclu}

In this work, we investigate the impact of galaxy interactions on the chemical properties of galaxies at high redshift. Using JWST observations from the PASSAGE Par028 field, we focus on galaxy pairs to study their gas-phase metallicities, metallicity gradients, and stellar masses, and to assess how interactions may perturb the mass–metallicity relation and internal chemical structure.

\begin{enumerate}
    \item We estimate the metallicity (\Z), metallicity gradient (\Zg), and stellar mass (\mstar) for 17 galaxies in candidate interacting systems in the JWST PASSAGE Par028 field. The sample spans redshifts $z \sim 1.7$--3.2, with stellar masses ranging from $\log\,(M_*/M_\odot) \sim 7.3$ to 10.4, \Z\ values between 7.63 and 8.70, and \Zg\ in the range of $-0.2$ to $+0.45$ dex/kpc.

    \item We find that while many interacting galaxies still broadly follow the expected mass--metallicity relations (MZRs), a non-negligible fraction, particularly at $z \sim 1.8$ and $z \sim 3$, exhibit metallicity offsets relative to the canonical MZR. 
    We further find tentative evidence that these metallicity offsets are linked to variations in the internal metallicity distribution. In particular, galaxies below the MZR preferentially exhibit flatter metallicity gradients, whereas systems with enhanced metallicities generally retain comparatively steeper or centrally concentrated metallicity distributions. The best-fit relation is
    $\nabla_{\rm [O/H]} = (-0.30\pm0.06) \, \Delta[12+\log(\mathrm{O/H})]_{\rm obs-MZR(z,\,M_{*})} + 0.04$.
    Although limited by the current sample size, this trend is consistent with a scenario in which galaxy interactions modify both the global metal enrichment and the radial distribution of metals through multiple gas dynamical processes. While tidal stripping and metal-poor gas inflows may dilute the central metallicity and flatten pre-existing gradients in some systems, interaction-driven central star formation may dominate in others, enhancing chemical enrichment in the nuclear regions and producing metallicities elevated above the canonical MZR.

    \item In our pair-selected sample, the mean gas-phase metallicity gradients are consistent with zero in the redshift bins centered at $z \sim 1.8$, 2.3, and 3.1. The individual gradients span approximately $-0.2$ to $+0.45$ dex kpc$^{-1}$. Compared with the heterogeneous literature compilation restricted to $0<z<3.2$, the sample appears to show a larger scatter, but the difference is not statistically significant. In the absence of a matched control sample, we cannot determine whether the observed distribution is driven by interactions, redshift evolution, or other differences between the samples.

    \item We investigate the correlation between \Zg and the stellar mass ratio within our pair-selected sample. For primary galaxies, we find a weak negative trend \Zg $\rm (dex/kpc)=(-0.13\pm 0.09)\log\,(mass\,ratio)$ (Spearman's $\rho=-0.16$, $p=0.655$), while secondary galaxies show a weak positive trend \Zg $\rm (dex/kpc)=(0.14\pm 0.14)\log\,(mass\,ratio)$ (Spearman's $\rho=0.21$, $p=0.662$). In both cases, the correlations are not statistically significant given the current sample size.

    \item Out of the five galaxy pairs in our sample for which metallicity gradients could be reliably measured for both members, four pairs exhibit gradients with opposite signs between the two galaxies. This suggests that merger-driven gas flows can affect the two galaxies differently, potentially due to variations in interaction stage, geometry, or localized inflows and outflows.

\end{enumerate}

\begin{acknowledgements}
We thank the anonymous referee for constructive comments that improved the quality of this paper.
This work is supported by the National Natural Science Foundation of China (grant 12373009), the National Key R\&D Program of China No.2025YFF0510603, the CAS Project for Young Scientists in Basic Research Grant No. YSBR-062, the China Manned Space Program with grant no. CMS-CSST-2025-A06, and the Fundamental Research Funds for the Central Universities. XW acknowledges the support by the Xiaomi Young Talents Program, and the work carried out, in part, at the Swinburne University of Technology, sponsored by the ACAMAR visiting fellowship.
We are very grateful to staff members at Space Telescope Science Institute who worked hard to schedule pure parallel opportunities for this project, especially to our Program Coordinator Shelly Meyett.
YD acknowledges the support by the National Key R\&D Program of China under grant No. 2022YFA1605300 and the National Nature Science Foundation of China (NSFC) grant No.12273051.
BV, AA, and PJW acknowledge support from the European Union – NextGenerationEU RFF M4C2 1.1 PRIN 2022 project 2022ZSL4BL INSIGHT, and from the INAF Mini Grant `1.05.24.07.01 RSN1: Spatially Resolved Near-IR Emission of Intermediate-Redshift Jellyfish Galaxies' (PI Watson).
AA and PJW acknowledge support from the INAF Large Grant 2022 “Extragalactic Surveys with JWST” (PI Pentericci).
AJB acknowledges funding from the "FirstGalaxies" Advanced Grant from the European Research Council (ERC) under the European Union’s Horizon 2020 research and innovation programme (Grant agreement No. 789056)

 The JWST/NIRISS data presented in this article were obtained from the Mikulski Archive for Space Telescopes (MAST) at the Space Telescope Science Institute. The specific PASSAGE Parallel-28 observations analyzed can be accessed via \dataset[doi:10.17909/h3ea-jd37]{https://doi.org/10.17909/h3ea-jd37}. The JWST/NIRCam imaging analyzed here consists of the COSMOS-Web DR1 mosaics described in Section~\ref{sec:datsam} and obtained from the COSMOS-Web public data release. The HST/ACS F814W imaging was obtained from the CANDELS high-level science products, available via \dataset[doi:10.17909/T94S3X]{https://doi.org/10.17909/T94S3X}.

\end{acknowledgements}

\clearpage
\begin{appendix}
\setcounter{figure}{0}
\renewcommand{\thefigure}{A\arabic{figure}}
\renewcommand{\theHfigure}{appendix.\arabic{figure}}
\section{Supplementary measurements and robustness tests}
\label{app:supplementary}
This appendix presents additional spatially resolved metallicity maps and radial profiles, followed by tests of the sensitivity of the measured gradients to emission-line S/N selection.

\subsection{Spatially resolved metallicity maps and radial profiles}
\label{app:profiles}
\begin{figure}[p]
    \centering
    \includegraphics[width=0.8\linewidth]{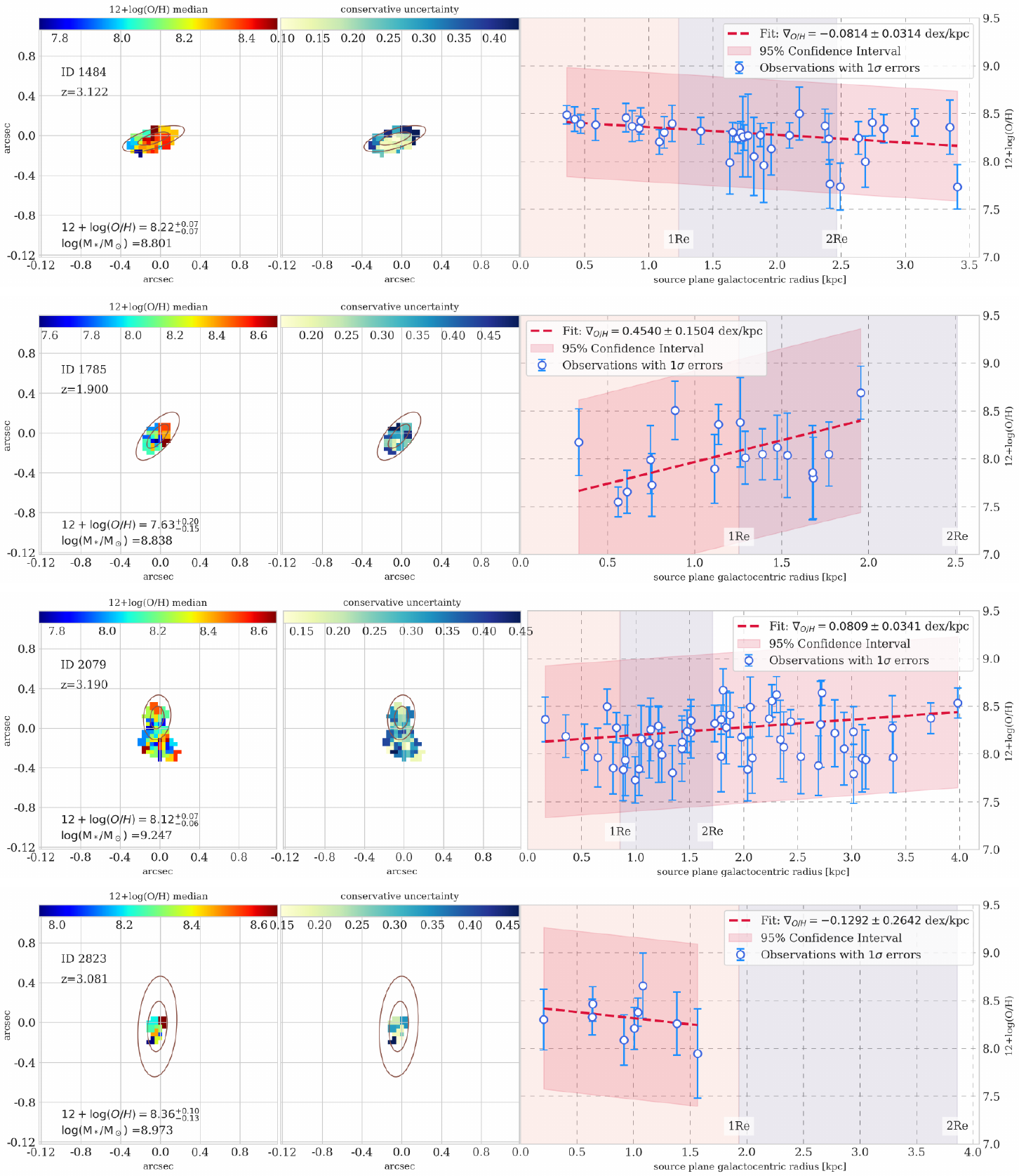}
    \caption{ Spatially resolved metallicity and radial profiles for systems in which only one member of the pair has sufficient emission-line signal for a metallicity-gradient measurement. For each galaxy, we present the 12+log (  O/H )  metallicity map (left), the conservative uncertainty map (middle), and the radial metallicity gradient (right). }
     \label{fig:appendix_single}
 \end{figure}

Figures~\ref{fig:appendix} and \ref{fig:appendix_single} present the metallicity maps, uncertainty maps, and radial profiles of galaxies not shown in the main-text example. Figure~\ref{fig:appendix}, continued across three pages, groups the profiles of paired components. Galaxy~1282 is displayed for completeness, but its two usable bins lie at nearly identical galactocentric radii and do not constrain a gradient; it is excluded from the quantitative analysis. Figure~\ref{fig:appendix_single} presents systems for which only one member has a usable metallicity-gradient measurement. Displaying a profile therefore does not necessarily imply that it enters the quantitative sample.

\begin{figure}[ht!]
    \centering
    \includegraphics[width=\linewidth,height=0.61\textheight,keepaspectratio]{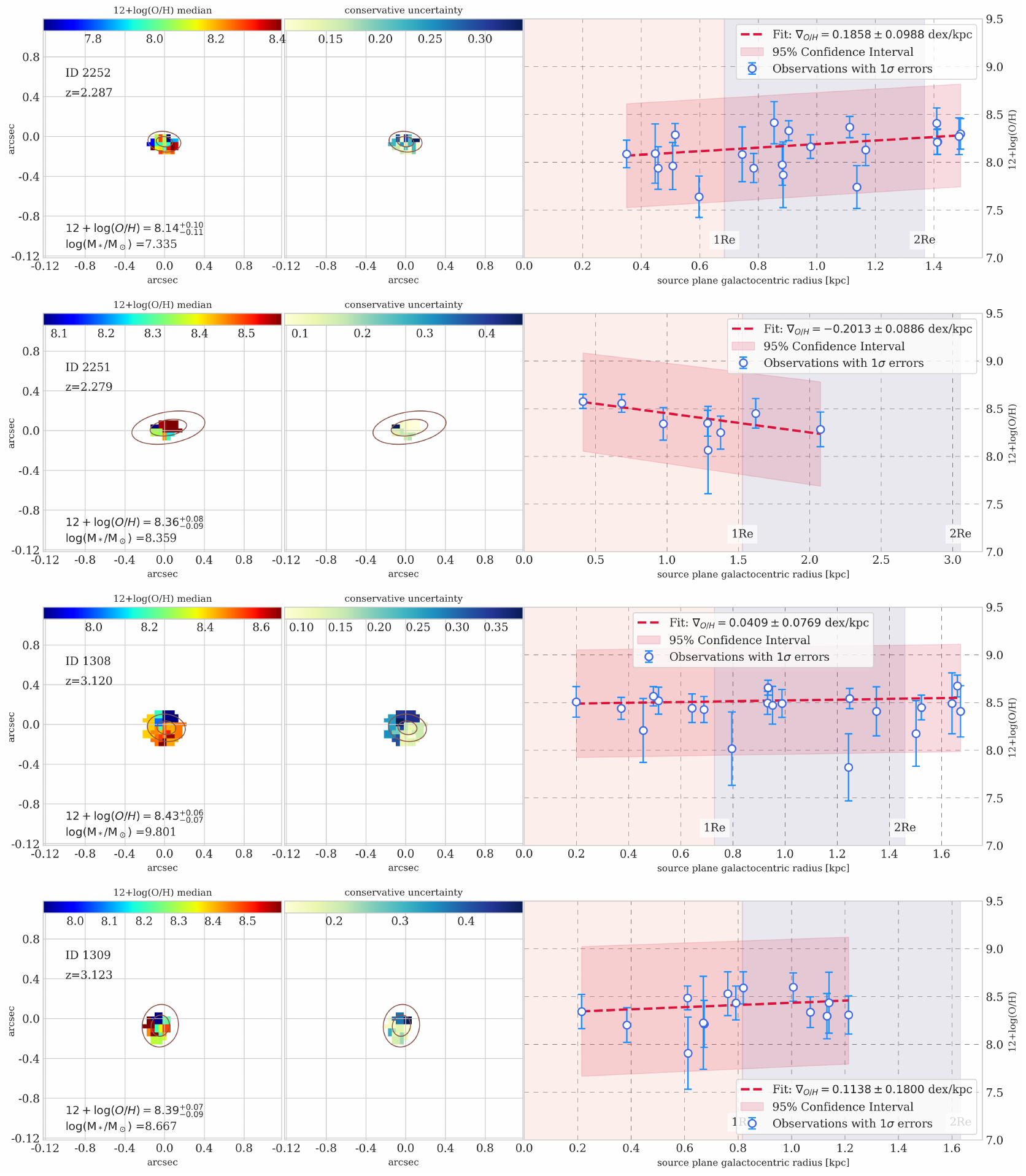}
    \caption{Spatially resolved metallicity and radial profiles of our sample galaxies. 
    For each galaxy, we present the 12+log(O/H) metallicity map (left), 
    the conservative uncertainty map (middle), and the radial metallicity gradient (right).}
    \label{fig:appendix}
\end{figure}

\begin{figure}[p]
    \centering
    \includegraphics[width=0.8\linewidth]{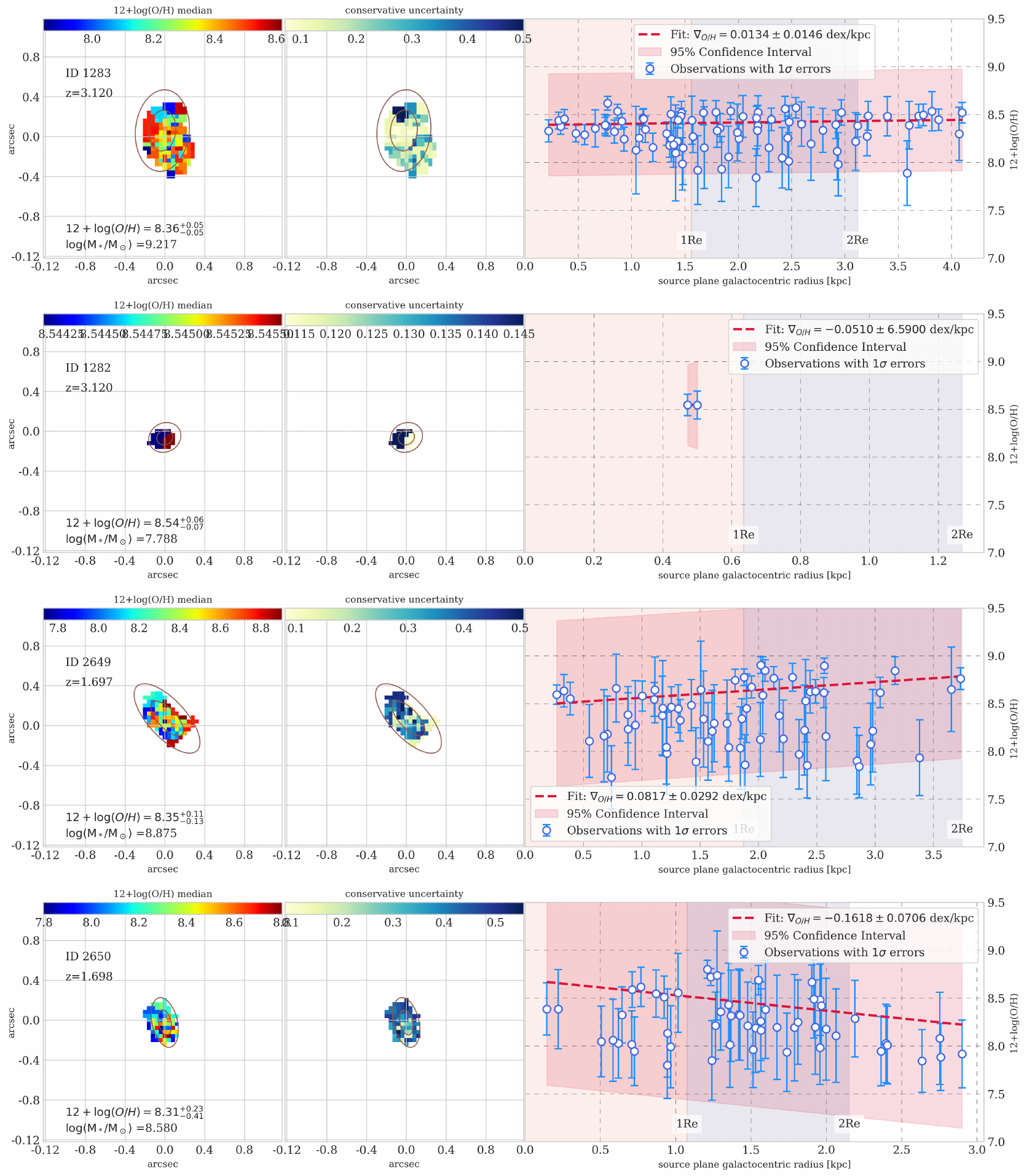}
     \addtocounter{figure}{-1}
     \caption{ Spatially resolved metallicity and radial profiles of the paired sample galaxies  (continued) . Galaxy~1282 is shown for completeness, but its gradient is not included in any quantitative analysis because only two usable spatial bins at nearly identical radii are available, leaving the slope unconstrained. }
\end{figure}

\begin{figure}[p]
    \centering
    \includegraphics[width=0.8\linewidth]{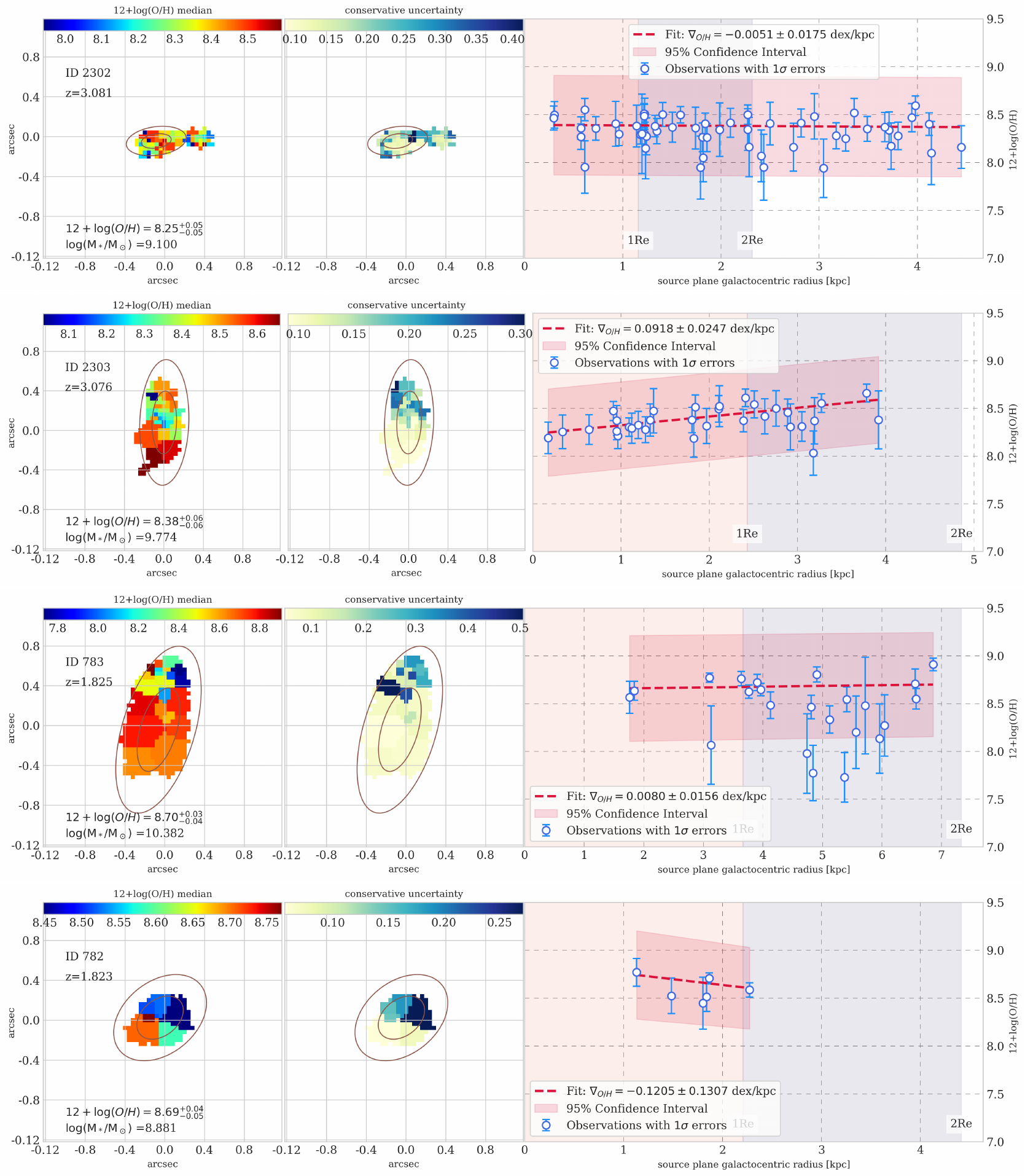}
     \addtocounter{figure}{-1}
     \caption{ Spatially resolved metallicity and radial profiles of the paired sample galaxies  (continued) . }
\end{figure}

\clearpage
\subsection{Sensitivity to emission-line S/N selection}
\label{app:snr_selection}

\begin{figure*}[ht!]
    \centering
    \includegraphics[width=\textwidth]{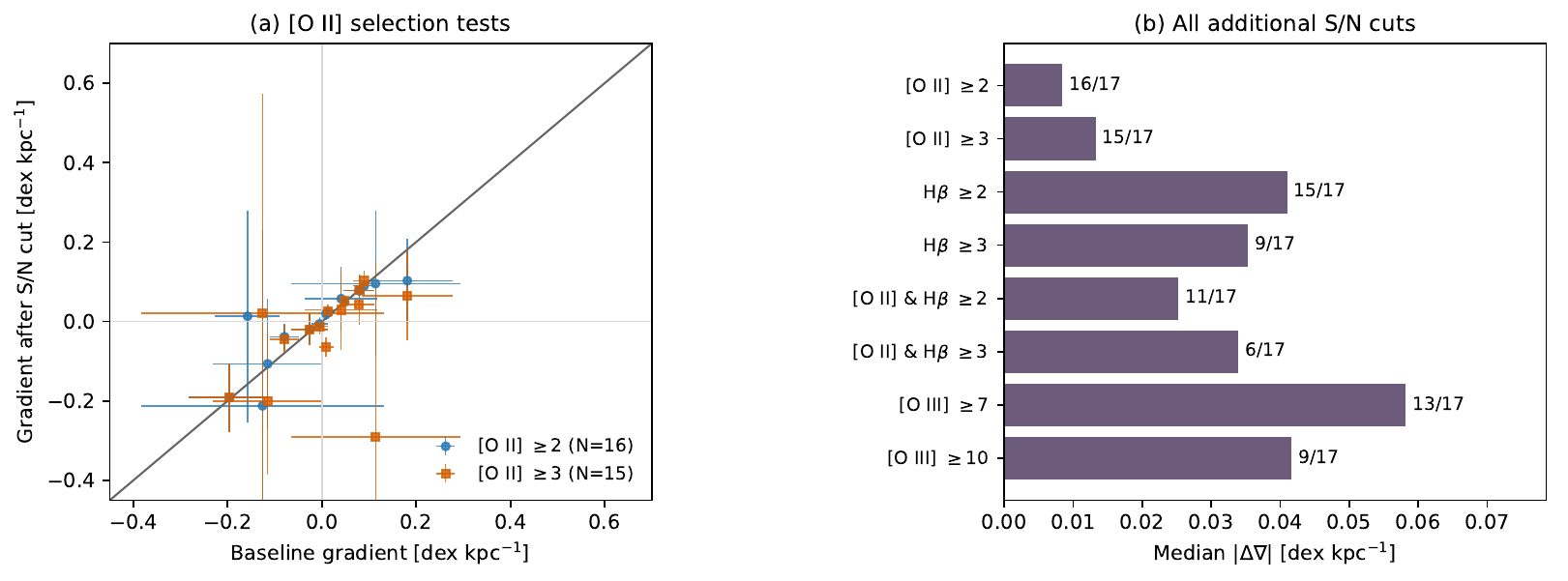}
    \caption{Sensitivity of radial metallicity gradients to additional emission-line S/N cuts for the 17-galaxy quantitative sample. Left: uniformly refitted baseline gradients versus gradients after requiring [O II] S/N of at least 2 or 3; the diagonal indicates equality and error bars show formal fitting uncertainties. Right: median absolute gradient changes for all tested cuts; labels give the number of measurable galaxies out of 17. The selection procedure and its limitations are described in Section~\ref{app:snr_selection}.}
    \label{fig:snr_robustness}
\end{figure*}

We test how additional emission-line S/N requirements affect the radial metallicity gradients of the 17 galaxies in the quantitative sample. Starting from the existing [O III]-selected Voronoi bins, we apply cuts on [O II], H$\beta$, their combination, and [O III], as summarized in Figure~\ref{fig:snr_robustness}. For each selection, we retain the original bin-level metallicity posteriors and galactocentric radii and repeat the inverse-variance-weighted linear fit using the surviving bins. We require at least three usable bins for a refit. These tests do not involve rebinning the maps or rerunning the metallicity inference. Galaxies 1282 and 1677 are excluded throughout.

Requiring [O II] S/N$\geq2$ leaves 16 measurable galaxies, with a median absolute gradient change of $0.008$ dex kpc$^{-1}$. Increasing the threshold to 3 leaves 15 galaxies and gives a median absolute change of $0.013$ dex kpc$^{-1}$. The corresponding median retained radial spans are 100\% and 93\% of the baseline spans. Figure~\ref{fig:snr_robustness} also shows that H$\beta$ and stricter [O III] cuts can produce larger changes. For example, H$\beta$ S/N$\geq3$ leaves nine measurable galaxies, with a median absolute change of $0.035$ dex kpc$^{-1}$ and a median retained radial span of 78\%. Requiring both [O II] and H$\beta$ S/N$\geq3$ leaves only six galaxies and reduces the median radial span to 54\%.

The [O II] cuts therefore give relatively small median changes, but individual gradients can remain sensitive to bin selection. More restrictive cuts alter both the sample membership and the radial coverage, so differences between the fits cannot be attributed solely to improved line significance. Moreover, the baseline and restricted fits share measurements and have correlated uncertainties; their formal error bars should not be treated as independent in a significance test. This exercise quantifies sensitivity to spatial-bin selection rather than independently validating the metallicity calibration or all low-S/N measurements.

\end{appendix}

  \bibliography{library_balanced_sfr}{}
 \bibliographystyle{aasjournal}

\end{document}